\pdfoutput=1

\documentclass[twocolumn]{emulateapj}

\usepackage{amsmath}
\usepackage{graphicx}
\usepackage{color}
\usepackage[colorlinks]{hyperref}
\hypersetup{
    colorlinks,	
    citecolor=blue,
}

\newcommand{\be}{\begin{eqnarray}}
\newcommand{\ee}{\end{eqnarray}}

\shorttitle{Testing GR with Galactic BHs}
\shortauthors{Tripathi et al.}

\begin{document}

\title{Testing General Relativity with \textsl{NuSTAR} data of Galactic Black Holes}

\author{Ashutosh~Tripathi\altaffilmark{1}, Yuexin~Zhang\altaffilmark{2}, Askar~B.~Abdikamalov\altaffilmark{1,3}, Dimitry~Ayzenberg\altaffilmark{4}, Cosimo~Bambi\altaffilmark{1,\dag}, Jiachen~Jiang\altaffilmark{5}, Honghui~Liu\altaffilmark{1}, and Menglei~Zhou\altaffilmark{6}}

\altaffiltext{1}{Center for Field Theory and Particle Physics and Department of Physics, 
Fudan University, 200438 Shanghai, China. \email[\dag E-mail: ]{bambi@fudan.edu.cn}}
\altaffiltext{2}{Kapteyn Astronomical Institute, University of Groningen, 9747 AD Groningen, The Netherlands}
\altaffiltext{3}{Ulugh Beg Astronomical Institute, Tashkent 100052, Uzbekistan}
\altaffiltext{4}{Theoretical Astrophysics, Eberhard-Karls Universit\"at T\"ubingen, D-72076 T\"ubingen, Germany}
\altaffiltext{5}{Department of Astronomy, Tsinghua University, 100084 Beijing, China}
\altaffiltext{6}{Institut f\"ur Astronomie und Astrophysik, Eberhard-Karls Universit\"at T\"ubingen, D-72076 T\"ubingen, Germany}

\begin{abstract}
Einstein's theory of General Relativity predicts that the spacetime metric around astrophysical black holes is described by the Kerr solution. In this work, we employ state-of-the-art relativistic reflection modeling to analyze a selected set of \textsl{NuSTAR} spectra of Galactic black holes to obtain the most robust and precise constraints on the Kerr black hole hypothesis possible today. Our constraints are much more stringent than those from other electromagnetic techniques, and with some sources we find stronger constraints than those currently available from gravitational waves. 
\end{abstract}


\section{Introduction}

The theory of General Relativity was proposed at the end of 1915~\citep{Einstein:1916vd} and so far has successfully passed a large number of observational tests. The theory has been extensively tested in the weak-field regime with experiments in the Solar System and observations of binary pulsars~\citep{Will:2014kxa}. The past five years have seen tremendous progress in the study of the strong-field regime and we can now test the predictions of General Relativity with gravitational waves~\citep[e.g.,][]{TheLIGOScientific:2016src,Yunes:2016jcc,LIGOScientific:2019fpa}, X-ray data~\citep[e.g.,][]{Cao:2017kdq,Tripathi:2018lhx,Tripathi:2020qco}, and mm and sub-mm Very Long Baseline Interferometry (VLBI) observations~\citep[e.g.,][]{Psaltis:2020lvx}.

In General Relativity and in the absence of exotic matter fields, the spacetime metric around astrophysical black holes should be described well by the Kerr solution~\citep{Kerr:1963ud}. This is the result of the black hole uniqueness theorems in General Relativity~\citep[see, e.g.,][]{2012LRR....15....7C} and of the fact that deviations induced by a possible accretion disk, nearby stars, or a black hole non-vanishing electric charge are normally negligible~\citep[see, e.g.,][]{2014PhRvD..89l7302B,2018AnP...53000430B}. However, macroscopic deviations from the Kerr metric are predicted in a number of scenarios with new physics, including models with macroscopic quantum gravity effects~\cite[e.g.,][]{Giddings:2017jts}, models with exotic matter fields~\cite[e.g.,][]{Herdeiro:2014goa}, or in the case General Relativity is not the correct theory of gravity~\citep[e.g.,][]{Kleihaus:2011tg}. Testing the Kerr metric around astrophysical black holes is thus a test of General Relativity in the strong field regime~\citep[but see][]{2008PhRvL.100i1101P}.

We note that electromagnetic and gravitational wave tests are, in general, complementary because they can probe different sectors of the theory~\citep{Bambi:2015kza}. Electromagnetic tests are more suitable to verify the interactions between the gravity and the matter sectors, including particle motion and non-gravitational interactions in a gravitational field~\citep{Will:2014kxa}. Gravitational wave tests can probe the gravitational sector itself in the strong and dynamical regime.

\begin{figure}[t]
\begin{center}
\includegraphics[width=8.5cm,trim={0cm 0cm 0cm 0cm},clip]{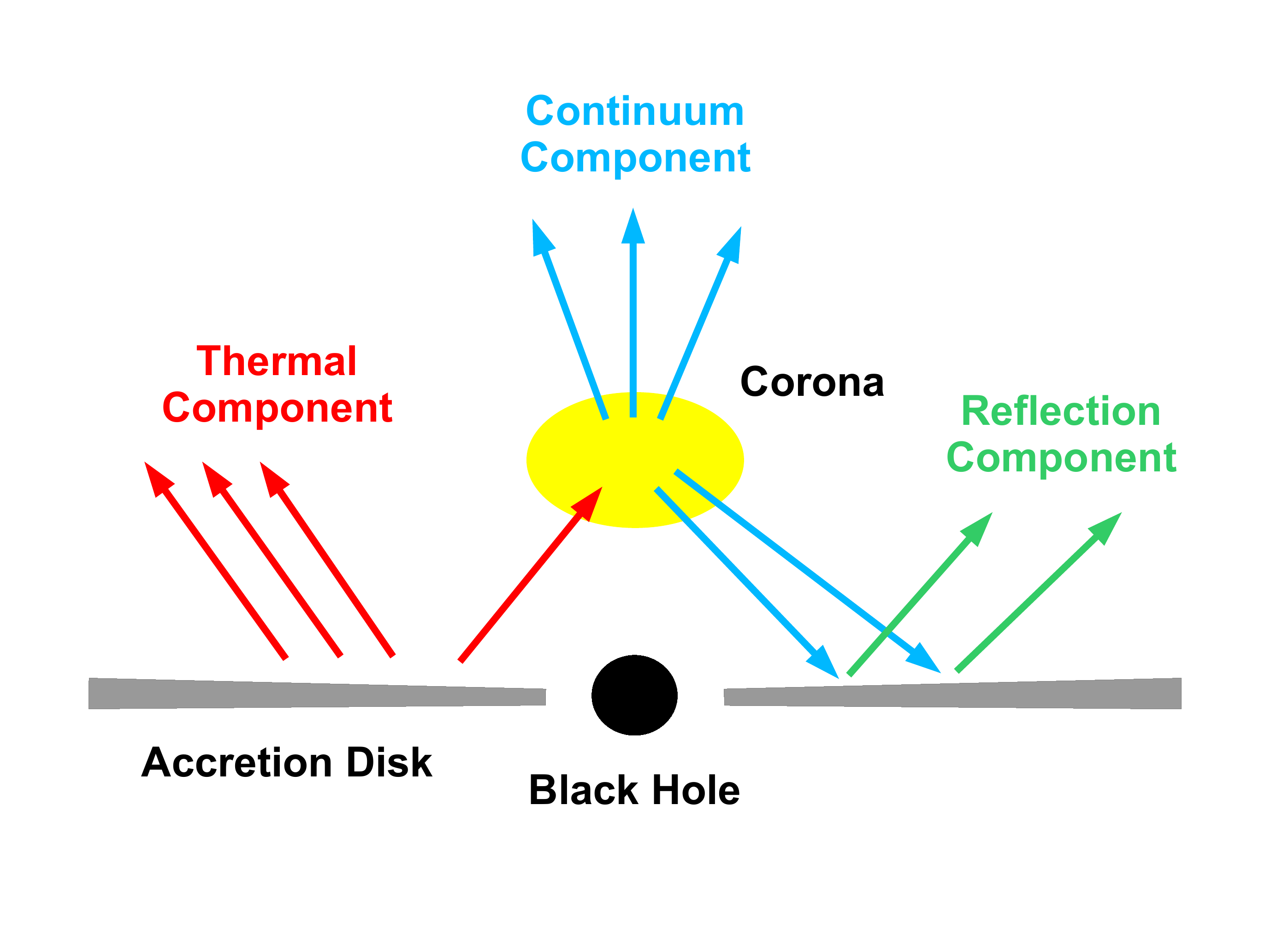}
\end{center}
\vspace{-1.3cm}
\caption{A black hole is accreting from a geometrically thin and optically thick disk. The disk has a multi-temperature blackbody-like spectrum. Thermal photons from the disk inverse Compton scatter off free electrons in the corona, producing the continuum component. The latter illuminates the disk, generating the reflection component. \label{f-corona}}
\end{figure}

The aim of the present work is to analyze the most suitable X-ray data of black holes to test the Kerr black hole hypothesis and to get the most robust constraints possible with the available observational data and theoretical models. Our astrophysical system is shown in Fig.~\ref{f-corona}~\citep[see, e.g.,][]{Bambi:2020jpe}. We have a black hole surrounded by a geometrically thin and optically thick cold (up to a few keV for stellar-mass black holes) accretion disk. Every point on the disk emits a blackbody-like spectrum and the whole disk has a multi-temperature blackbody-like spectrum. Thermal photons from the disk inverse Compton scatter off free electrons in the so-called corona, which is a hotter ($\sim$100~keV) gas close to the black hole and the inner part of the accretion disk. The Comptonized photons have a power-law spectrum with a high-energy cutoff and can illuminate the disk, producing the reflection spectrum. The latter is characterized by some fluorescent emission lines in the soft X-ray band, notably the iron K$\alpha$ complex peaked at $\sim$6~keV, and by a Compton hump peaked at 20-30~keV~\citep{Ross:2005dm,Garcia:2010iz}. In the presence of high-quality data and employing the correct astrophysical model, the analysis of these reflection features emitted from the inner part of the accretion disk can be a powerful tool to probe the strong gravity region near the black hole event horizon~\citep{2019NatAs...3...41R,Bambi:2020jpe}.

The paper is organized as follows. In Section~\ref{s:obs}, we present our selection of sources and observations. Section~\ref{s:ana} is devoted to the data reduction and analysis. In Section~\ref{s:dis}, we discuss our results and we compare our constraints with those obtained from mm and sub-mm VLBI observations and from gravitational waves. In our study, we employ the Johannsen metric with the deformation parameter $\alpha_{13}$~\citep{Johannsen:2015pca}; its expression is reported in Appendix~\ref{a:johannsen}. In principle, we could use any other stationary, axisymmetric, and asymptotically flat non-Kerr black hole metric to constrain the associated deformation parameter and quantify possible deviations from the Kerr geometry. However, rotating black hole solutions beyond General Relativity are normally known either in numerical form or in an expansion on the spin parameter. Since our model currently works only with analytic metrics and we want to analyze the spectra of fast-rotating objects, with spin parameter close to 1, here we have employed a phenomenological metric. We also note that so far our efforts have been mainly focused to improve the astrophysical model to limit/understand the systematic uncertainties, thus devoting less attention to the capabilities of X-ray reflection spectroscopy of constraining specific gravity theories.


\section{Selection of the sources/observations} \label{s:obs}

The study of relativistic reflection features can be used to test the Kerr metric around either stellar-mass black holes in X-ray binaries and supermassive black holes in active galactic nuclei (AGN). We note that in some theoretical models deviations from the Kerr metric can be more easily detected in one of the two object classes than in the other one. For example, in higher-curvature theories, such as dynamical Chern-Simons and Einstein-dilaton-Gauss-Bonnet, observations of stellar-mass black holes normally lead to much stronger constraints than those of supermassive black holes, because the curvature at the black hole horizon is inversely proportional to the square of the black hole mass~\citep{Yagi:2016jml}. We also note that tests with X-ray binaries have some advantages over AGN. The sources are brighter, and so have better statistics. The spectra of AGN often present warm absorbers, which may introduce astrophysical uncertainties in the measurement. In this study, we thus focus on stellar-mass black holes.

\textsl{NuSTAR}~\citep{Harrison:2013md} is the most suitable X-ray mission for the study of relativistic reflection features of black hole binaries. The data are not affected by pile-up from the brightness of the sources and cover a broad energy band from 3 to 79~keV, so they permit us to analyze both the iron line and the Compton hump. Our starting point is thus the list of all published spin measurements of stellar-mass black holes with \textsl{NuSTAR} data (assuming the Kerr metric) obtained by studying the reflection features of the source. This list is shown in Tab.~\ref{t-bhb} and we see fourteen objects.

\begin{table*}
    \centering
    \renewcommand\arraystretch{1.5}{
    \begin{tabular}{ccccc}
    \hline\hline
    Source & $a_*$ & Mission(s) & State & Reference \\
    \hline\hline
    4U~1630--472 & $0.985_{-0.014}^{+0.005}$ & \textsl{NuSTAR} & Intermediate & \citet{King:2014sja} \\
    \hline    
    Cyg~X-1& $>0.83$ & \textsl{NuSTAR}+\textsl{Suzaku} & Soft & \citet{Tomsick:2013nua} \\
    & $>0.97$ & \textsl{NuSTAR}+\textsl{Suzaku} & Hard & \citet{Parker:2015fja} \\
    & $0.93 \sim 0.96$ & \textsl{NuSTAR} & Soft & \citet{Walton:2016hvd} \\
    \hline
    EXO 1846--031 & $0.997_{-0.002}^{+0.001}$ & \textsl{NuSTAR} & Hard intermediate & \citet{Draghis:2020ukh} \\
    \hline    
    GRS~1716--249 & $>0.92$ & \textsl{NuSTAR}+\textsl{Swift} & Hard Intermediate & \citet{Tao:2019yhu} \\
    \hline
    GRS~1739--278 & $0.8 \pm 0.2$ & \textsl{NuSTAR} & Low/Hard & \citet{Miller:2014sla} \\
    \hline
    GRS~1915+105 & $0.98 \pm 0.01$ & \textsl{NuSTAR} & Low/Hard & \citet{Miller:2013rca} \\
    \hline    
    GS~1354--645 & $>0.98$ & \textsl{NuSTAR} & Hard & \citet{El-Batal:2016wmk} \\
    \hline
    GX~339--4 & $0.95_{-0.08}^{+0.02}$ & \textsl{NuSTAR}+\textsl{Swift} & Very High & \citet{Parker:2016ltr} \\
    \hline
    MAXI~J1535--571 & $>0.84$ & \textsl{NuSTAR} & Hard & \citet{Xu:2017yrm} \\
    \hline
    MAXI~J1631--479 & $>0.94$ & \textsl{NuSTAR} & Soft & \citet{Xu:2020vil} \\
    \hline    
    Swift~J1658.2--4242 & $>0.96$ & \textsl{NuSTAR}+\textsl{Swift} & Hard & \citet{Xu:2018lfo} \\
    \hline
    Swift~J174540.2--290037 & $0.92_{-0.07}^{+0.05}$ & \textsl{Chandra}+\textsl{NuSTAR} & Hard & \citet{Mori:2019iwz} \\    
    \hline
    Swift~J174540.7--290015 & $0.94_{-0.10}^{+0.03}$ & \textsl{Chandra}+\textsl{NuSTAR} & Soft & \citet{Mori:2019iwz} \\
    \hline
    V404~Cyg & $>0.92$ & \textsl{NuSTAR} & Hard & \citet{Walton:2016fso}\\
    \hline\hline
\end{tabular} }
\vspace{0.2cm}
\caption{\rm List of all published spin measurements of stellar-mass black holes obtained from \textsl{NuSTAR} data and from the analysis of relativistic reflection features of the sources. The fourth column refers to source state during the observation used for the spin measurement. All these spectral states are explicitly mentioned in reference papers in the last column except for V404~Cyg. The spectra of V404 Cyg in~\citet{Walton:2016fso} exhibit a very hard power-law ($\Gamma \approx 1.4$) and a weak disk blackbody component, so they can be classified in the hard state. \label{t-bhb}}
\end{table*}

We want to now select the best sources and observations to get robust tests of the Kerr metric, as some sources/data are problematic and our result could be affected by systematic uncertainties not under control. Cyg~X-1 is a complicated source because it is a high mass X-ray binary and the spectrum of the accretion disk of the black hole is strongly absorbed by the wind of the companion star. In the observation of GRS~1716--249 it is unclear whether the disk extends up to the innermost stable circular orbit (ISCO) or is truncated at a somewhat larger radius~\citep{Jiang:2019gfa}. GRS~1915+105 is another well-known complicated source, where outflows from the accretion disk can absorb the X-ray radiation emitted from the region near the black hole. The observation of MAXI~J1535--571 is affected by dust scattering. MAXI~J1631--497 was observed in the soft state, where our reflection model for cold disks may have some problem, and the reflection spectrum is strongly affected by a variable and extremely fast disk wind. Swift~J174540.2 and Swift~J174540.7 are two X-ray transients within 1~pc from Sgr~A*, which makes a precise measurement of these sources quite challenging (it is even under debate whether these sources are indeed black hole binaries or not). V404~Cyg is another very complicated source characterized by strong and extremely variable absorption. Eventually, we remain with six sources more suitable (even if not all ideal) for tests of the Kerr black hole hypothesis: 4U~1630--472, EXO~1846--031, GRS~1739--278, GS~1354--645, GX~339--4, and Swift~J1658--4242.


\section{Data reduction and analysis} \label{s:ana}

\subsection{Data reduction}

\begin{table*}
 \centering
 \renewcommand\arraystretch{1.5}
\begin{tabular}{ccccc}
\hline\hline
\hspace{0.1cm} Source \hspace{0.1cm} & \hspace{0.1cm} Observation ID \hspace{0.1cm} & \hspace{0.1cm} Observation Date \hspace{0.1cm} & \hspace{0.1cm} Exposure (ks) \hspace{0.1cm} & \hspace{0.1cm} Counts [s$^{-1}$] \hspace{0.1cm} \\
\hline\hline
4U~1630--472              & 40014009001 & 2013 May 9    & 14.6  &  77.5\\
\hline
EXO~1846--031              & 90501334002 & 2019 August 3 & 22.2  & 148.7 \\
\hline
GRS~1739--278             & 80002018002 & 2014 March 26 & 29.7  & 127.8 \\
\hline
GS~1354--645              & 90101006004 & 2015 July 11  & 30.0  & 51.8\\
\hline
GX~339--4& 80001015003 & 2015 March 11 & 30.0  & 208.5\\
                         & 00081429002 & 2015 March 11 & 1.9   & 35.4 \\
                         \hline
Swift~J1658--4242& 90401307002 & 2018 February 16 & 33.3  & 34.5\\
                         & 00810300002 & 2018 February 16 & 3.0   & 4.9 \\
                         \hline\hline
\end{tabular}
 \caption{\rm Summary of the sources and the observations analyzed in the present work. \label{t-obs}}
\end{table*}

The observations of the six selected sources are listed in Tab.~\ref{t-obs}. In the case of GX~339--4 and Swift~J1658--4242, the first observation ID refers to the \textsl{NuSTAR} observation, the second one to the \textsl{Swift} observation. We note that these data of GS~1354--645 and GX~339--4 were already analyzed with an earlier version of our reflection model in \citet{Xu:2018lom} and \citet{Tripathi:2020dni}, respectively, but we repeat the analysis here with the latest version of our model (which, among other things, is more accurate for high inclination angles) in order to be able to consistently combine the measurements from the six sources together.

The raw data of the \textsl{NuSTAR} detectors FPMA and FPMB are processed to get cleaned events using the script {\tt nupipeline} of NuSTAR data analysis Software (NustarDAS) v2.0.0, which is a part of spectral analysis HEASOFT v6.28. We use the latest calibration database CALDB v20200912. The source region, which varies from source to source, is selected such that 90\% of its photons lie within. The background region, which has a similar size to the source region, is selected far from the source to avoid inclusion of source photons. The spectrum, ancillary, and response files are created using the module {\tt nuproducts} of NuSTARDAS.

For the XRT/\textsl{Swift} data, the raw data are processed to cleaned event files using the {\tt xrtpipeline} script of XRT/\textsl{Swift} data analysis Software (SWXRTDAS) v3.5.0 and CALDB v20200724. The source and background regions are selected in a similar way explained above for the \textsl{NuSTAR} data. The cleaned event files are then used to extract source and background spectra using {\tt xselect} v2.4g. Ancillary files are generated using {\tt xrtmkarf} and response matrices are taken from CALDB.

Finally, the source spectrum is grouped depending on the brightness of the source so that $\chi^2$ statistics can be applied. XSPEC v12.11.1 is used for spectral analysis and, in what follows, all the errors quoted are at 1-$\sigma$ unless stated otherwise.

\subsection{Data analysis}

\subsubsection{Individual measurements}

\begin{figure*}
\begin{center}
\includegraphics[width=8.5cm,trim={1.5cm 0.5cm 3.5cm 18.0cm},clip]{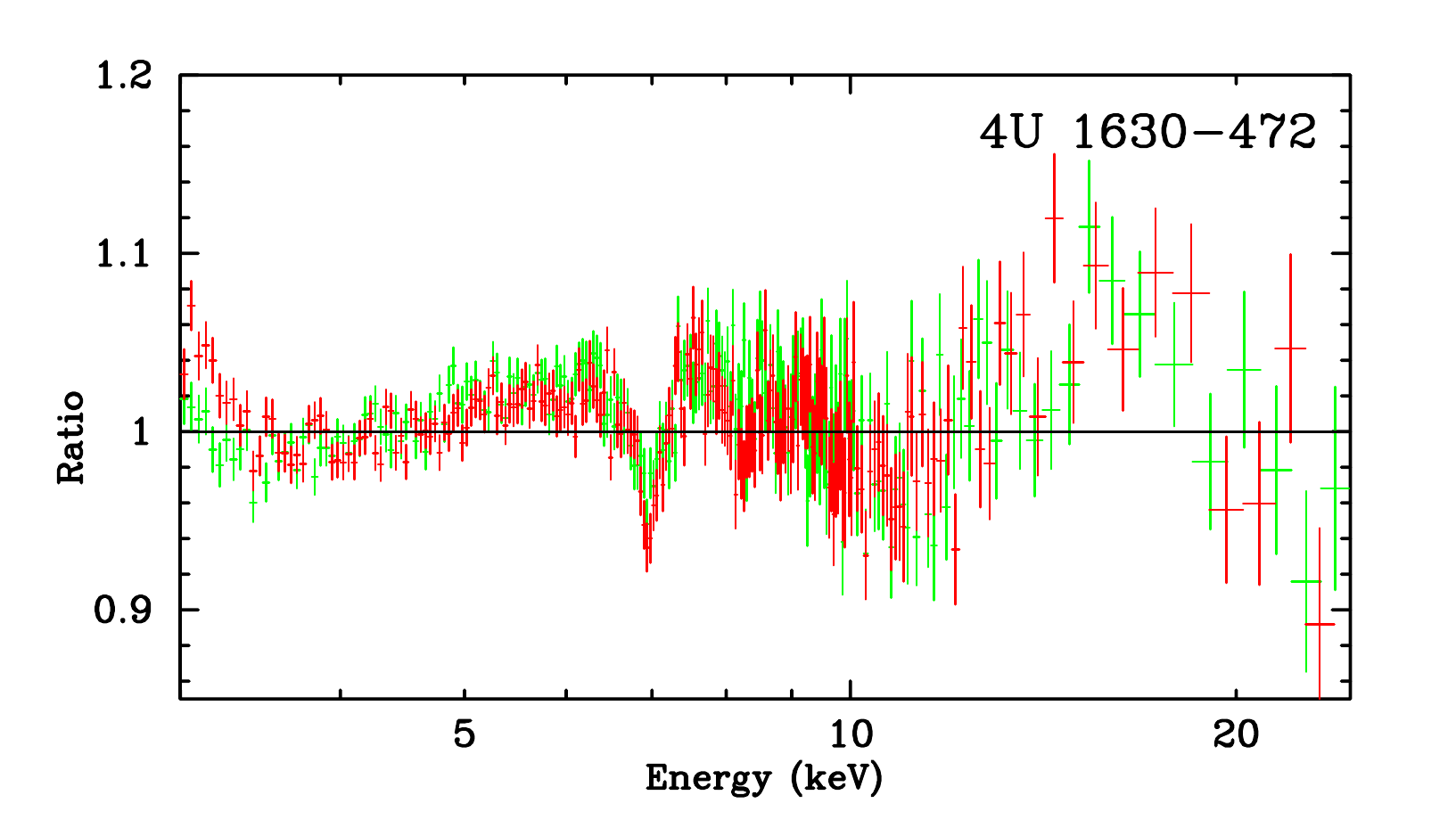}
\includegraphics[width=8.5cm,trim={1.5cm 0.5cm 3.5cm 18.0cm},clip]{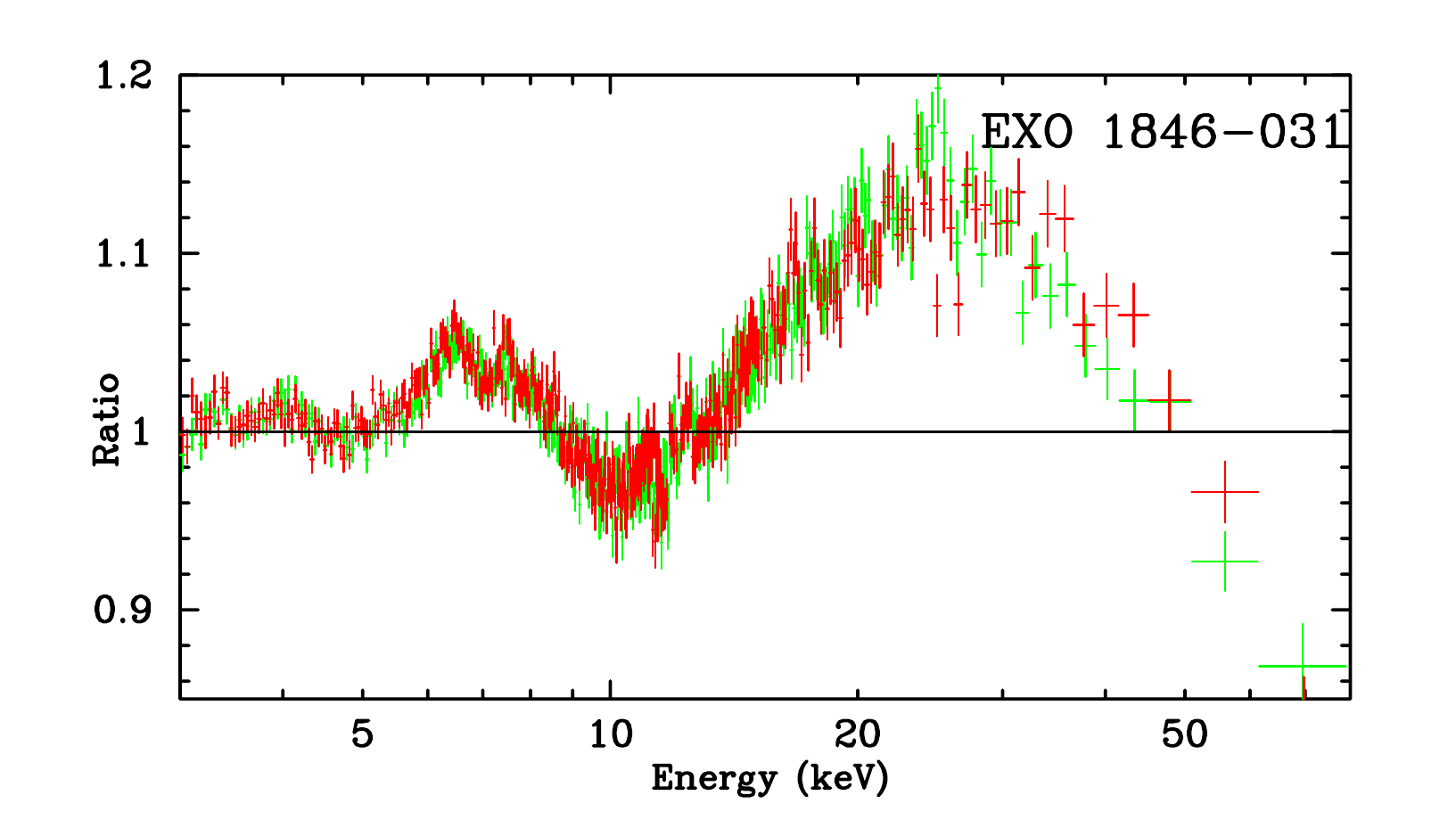} \\
\includegraphics[width=8.5cm,trim={1.5cm 0.5cm 3.5cm 18.0cm},clip]{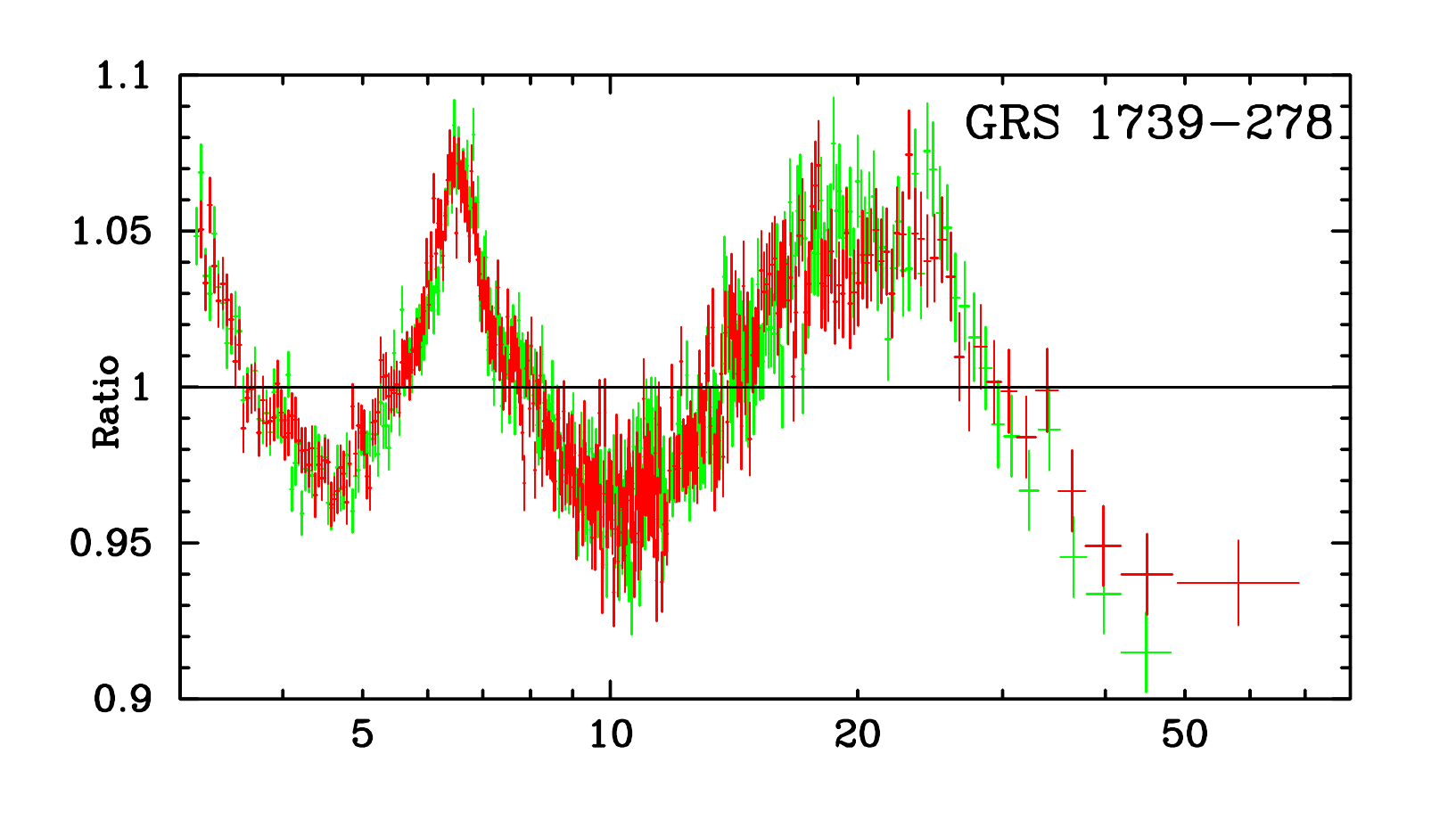}
\includegraphics[width=8.5cm,trim={1.5cm 0.5cm 3.5cm 18.0cm},clip]{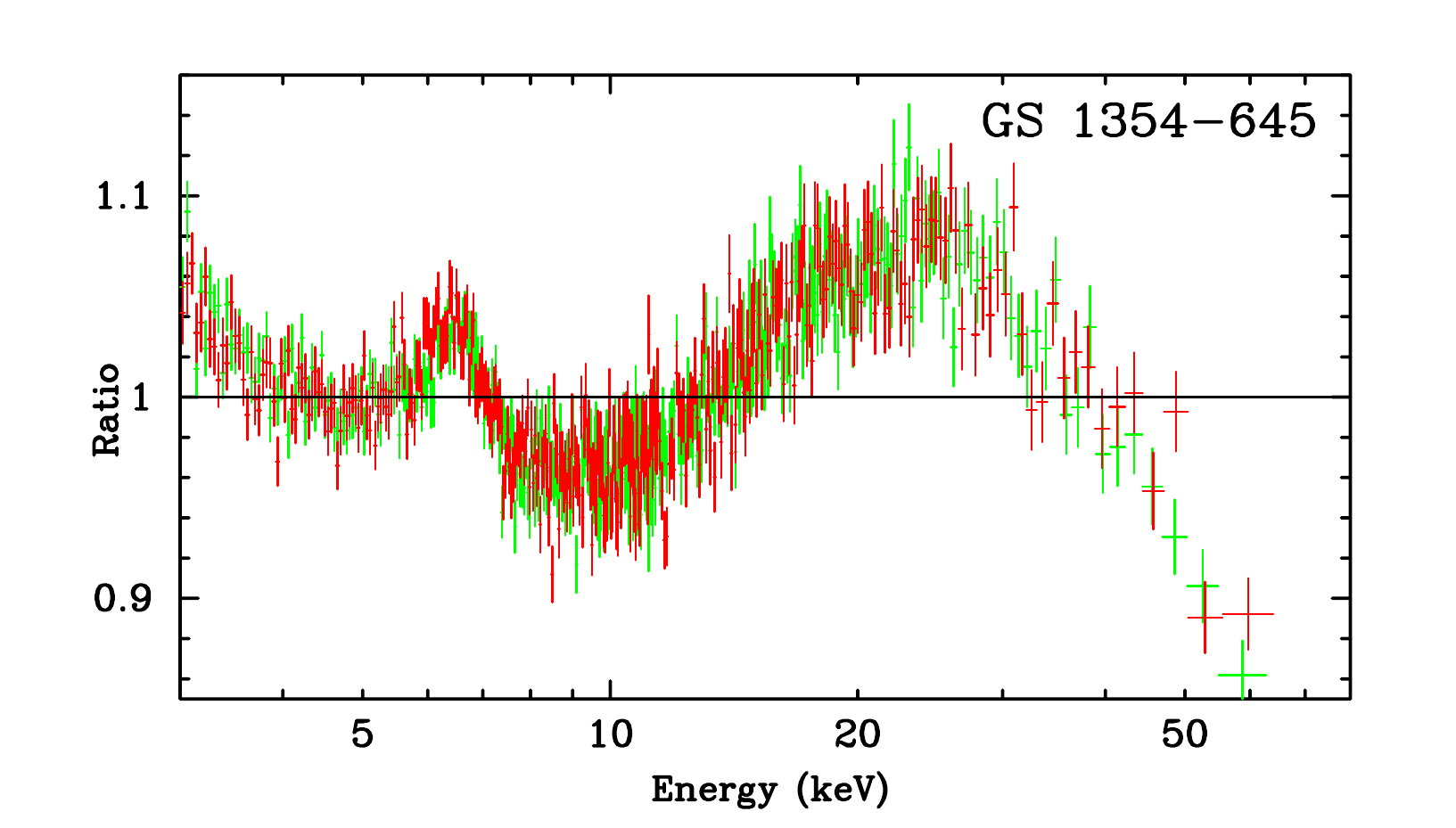} \\
\includegraphics[width=8.5cm,trim={1.5cm 0.5cm 3.5cm 18.0cm},clip]{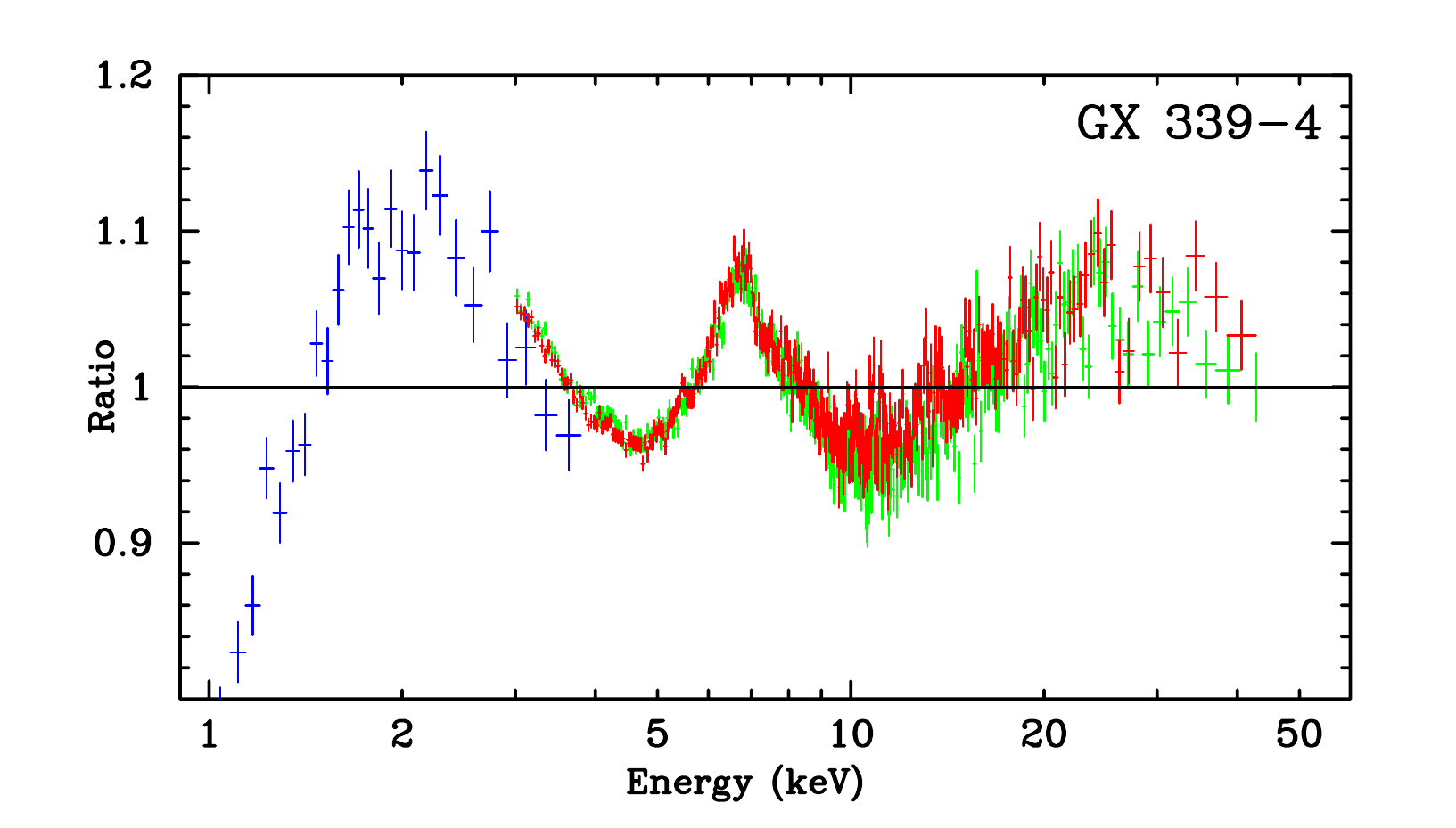}
\includegraphics[width=8.5cm,trim={1.5cm 0.5cm 3.5cm 18.0cm},clip]{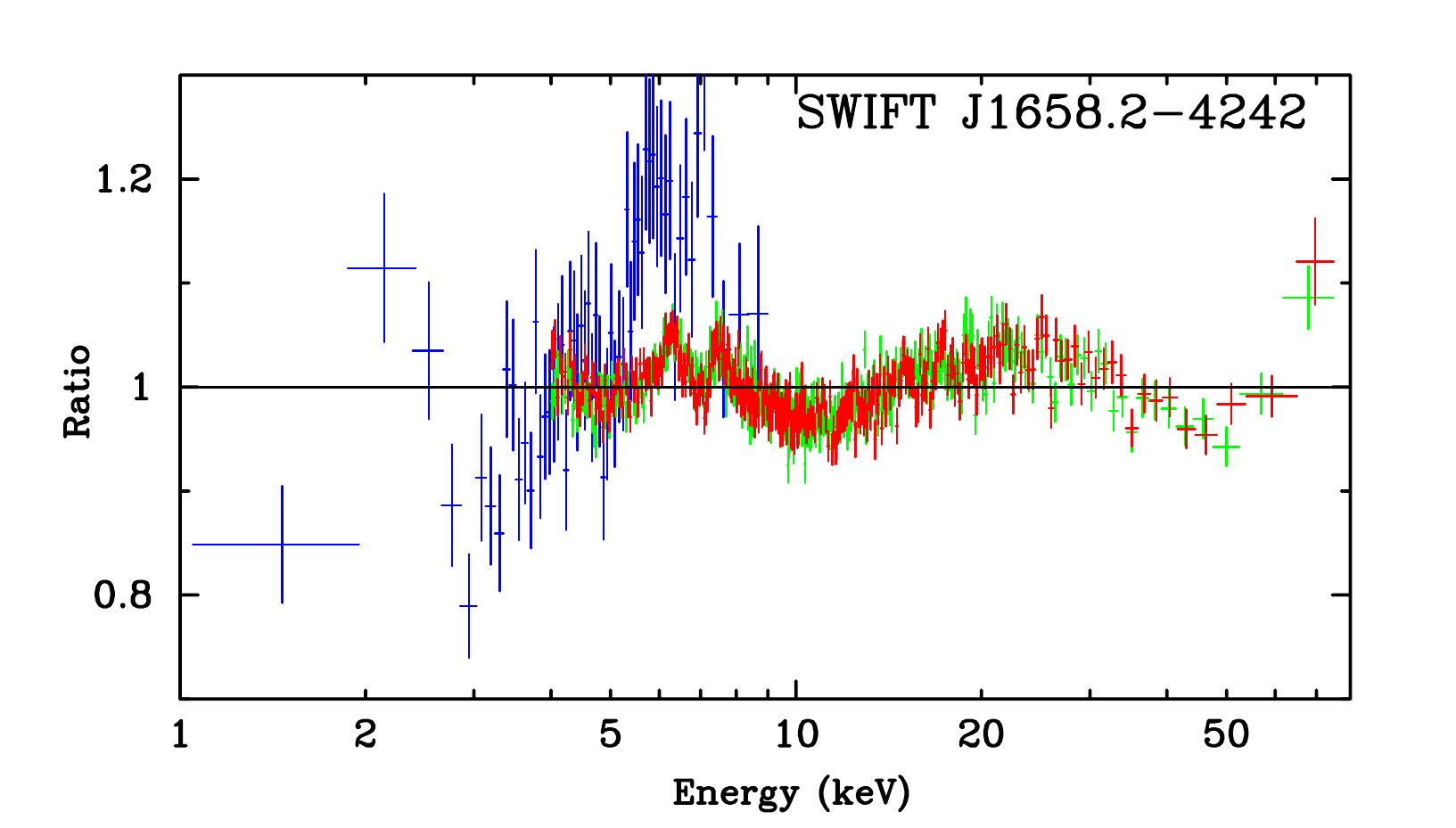}
\end{center}
\vspace{-0.5cm}
\caption{Data to best-fit model ratio for an absorbed power-law (an absorbed power-law + disk blackbody spectrum in the case of GX~339--4) for the spectra analyzed in this work. Green crosses are for FPMA/\textsl{NuSTAR} data, red crosses are for FPMB/\textsl{NuSTAR} data, and blue crosses are for XRT/\textsl{Swift} data. \label{f-ratio}}
\end{figure*}

First, we analyze every source separately. We start by fitting the data with an absorbed power-law (an absorbed power-law + disk blackbody spectrum in the case of GX~339--4) and we show the data to best-fit model ratios for the six sources in Fig.~\ref{f-ratio}. For every source, we clearly see a broad iron line and a Compton hump, indicating that we have a relativistically blurred reflection spectrum. In the case of 4U~1630--472, there is strong absorption at the iron line, but we can still determine that a broad iron line is there.

For the analysis of the reflection spectrum, we employ {\tt relxill\_nk}~\citep{Bambi:2016sac,Abdikamalov:2019yrr}, which is an extension of the {\tt relxill} package~\citep{Dauser:2013xv,Garcia:2013lxa} to non-Kerr spacetimes. The model and its parameters are briefly reviewed in Appendix~\ref{a:relxillnk}. We use the latest version of {\tt relxill\_nk}, corresponding to v1.4.0. In the case of GX~339--4, the spectrum presents a prominent thermal component of the accretion disk, and thus we also use {\tt nkbb}~\citep{Zhou:2019fcg}. {\tt relxill\_nk} and {\tt nkbb} are the state-of-the-art in, respectively, reflection and thermal models in non-Kerr spacetimes. In our analysis, we assume that the spacetime around these black holes can be described by the Johannsen metric~\citep{Johannsen:2015pca} with the possible non-vanishing deformation parameter $\alpha_{13}$, while all other deformation parameters are set to zero (see Appendix~\ref{a:johannsen} for more details). $\alpha_{13}$ is thus used to quantify possible deviations from the Kerr background, which is recovered for $\alpha_{13} = 0$. From the analysis of the \textsl{NuSTAR} data of these six sources we want to estimate $\alpha_{13}$ and see whether it is consistent with zero, as is necessary to recover the Kerr metric and is predicted by General Relativity.

We proceed to find the best model for every source. Note that we cannot directly fit the data with the model published in literature for the spin measurement in the Kerr metric because, in principle, deviations from the Kerr background may be able to explain the data with different spectral components. However, we note that eventually we use the best models already used for the spin measurements of these sources, which can be explained with the fact that deviations from the Kerr background due to a non-vanishing $\alpha_{13}$ have their own impact on the spectrum and do not affect the choice of the final model (but such a conclusion may not hold for other deformations from the Kerr metric). Eventually we fit the data with the models listed in Tab.~\ref{tab-m}.

\begin{table*}
\centering
{\renewcommand{\arraystretch}{1.5}
\begin{tabular}{ll}
\hline\hline
Source \hspace{2.0cm} &  Model \\
\hline\hline
4U~1630--472 &
{\tt tbabs$\times$xstar$\times$relxilllpCp\_nk} \\
\hline
EXO~1846--031 &
{\tt tbabs$\times$(diskbb + relxill\_nk + gaussian)} \\
\hline
GRS~1739--278 &
{\tt tbabs$\times$relxill\_nk} \\
\hline
GS~1354--645 &
{\tt tbabs$\times$relxillCp\_nk} \\
\hline
GX~339--4 &
{\tt tbabs$\times$(nkbb + comptt + relxilllp\_nk)} \\
\hline
Swift~J1658.2--4242 &
{\tt tbnew$\times$xstar$\times$(nthComp + relxilllpCp\_nk + gaussian)} \hspace{0.5cm} \\
\hline\hline
\end{tabular}
}
\caption{\rm List of the models used for every source. \label{tab-m}}
\end{table*}

\begin{figure*}[t]
\begin{center}
\includegraphics[width=8.5cm,trim={2.0cm 0.5cm 3.0cm 17.5cm},clip]{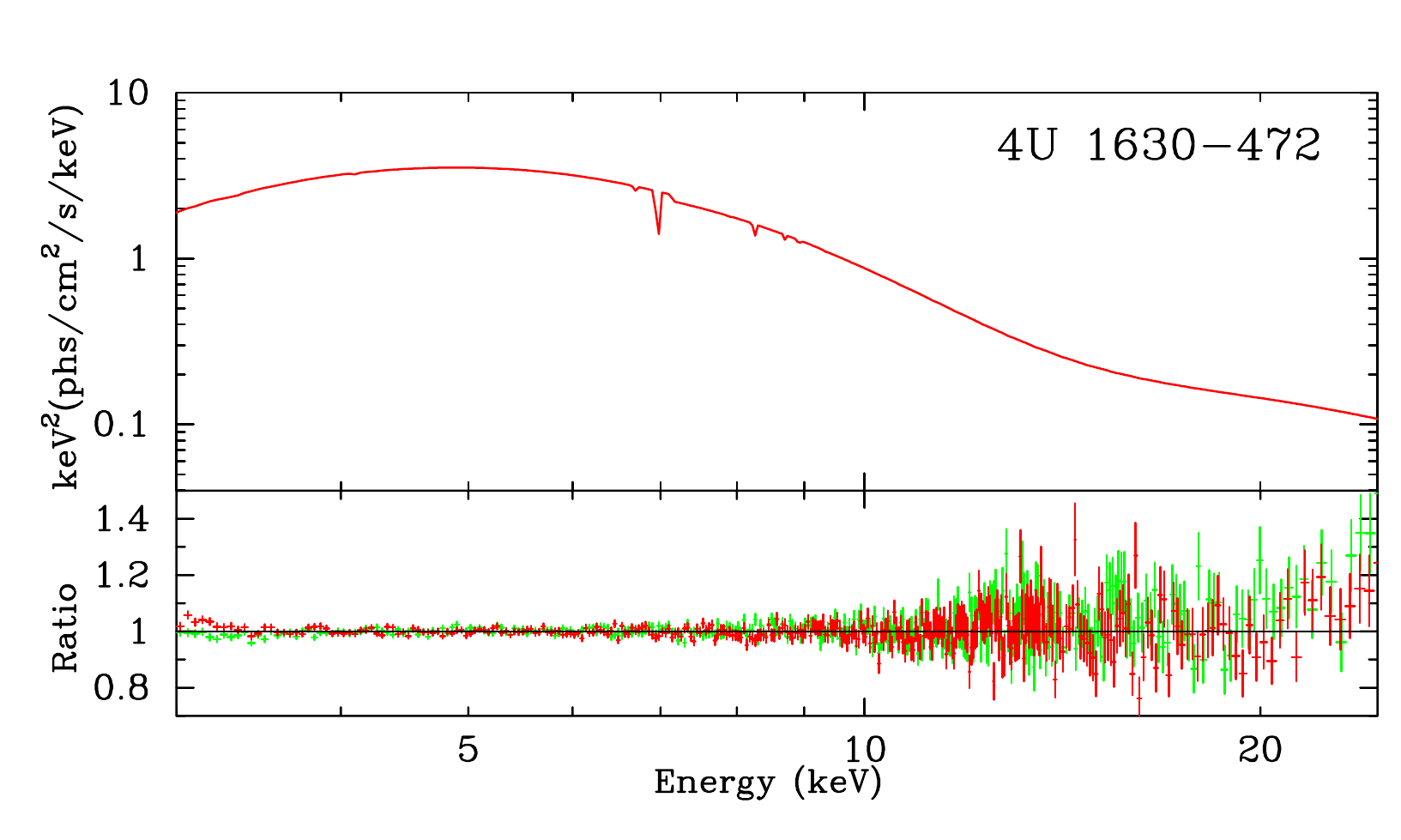}
\hspace{0.25cm}
\includegraphics[width=8.5cm,trim={2.0cm 0.5cm 3.0cm 17.5cm},clip]{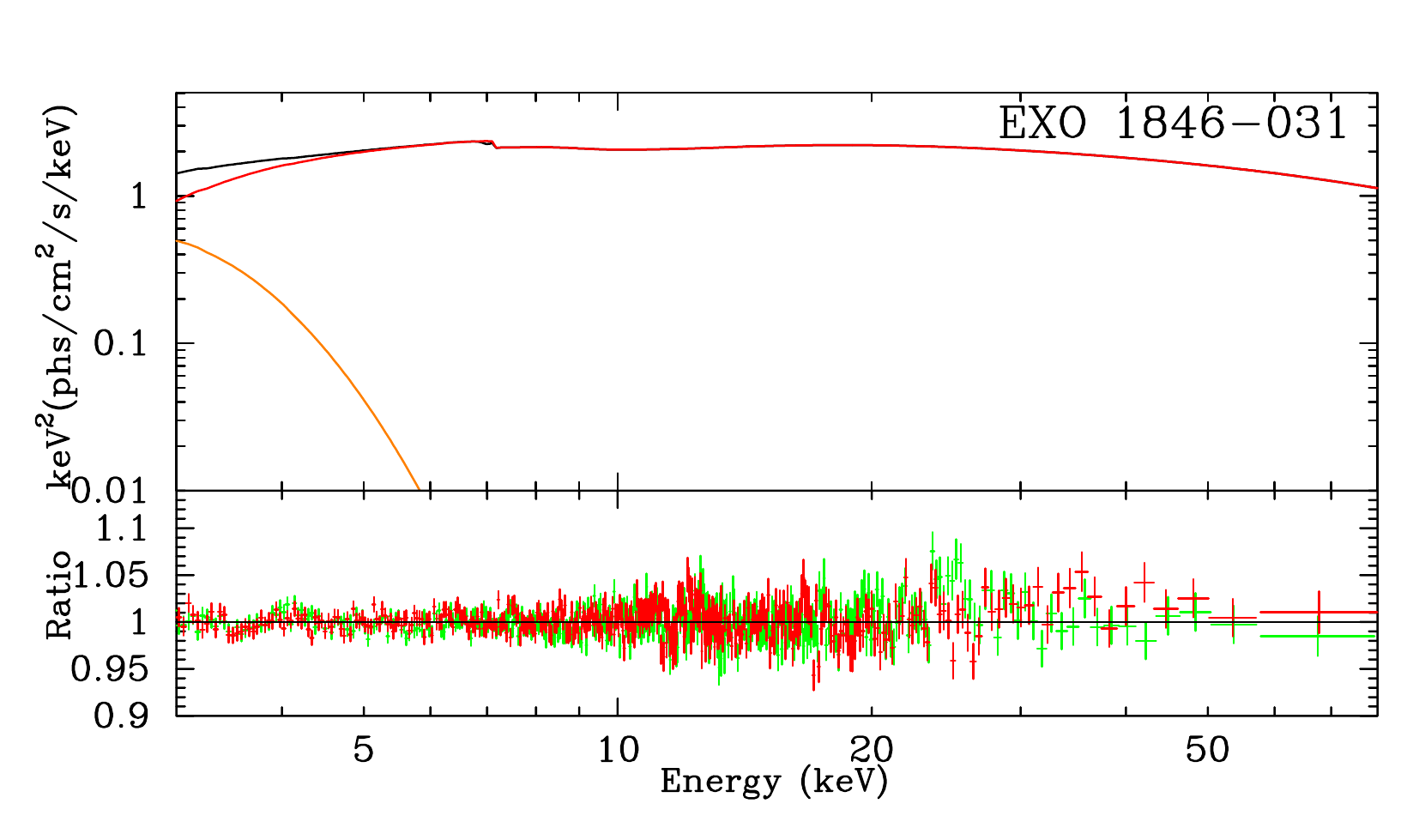} \\
\includegraphics[width=8.5cm,trim={2.0cm 0.5cm 3.0cm 17.5cm},clip]{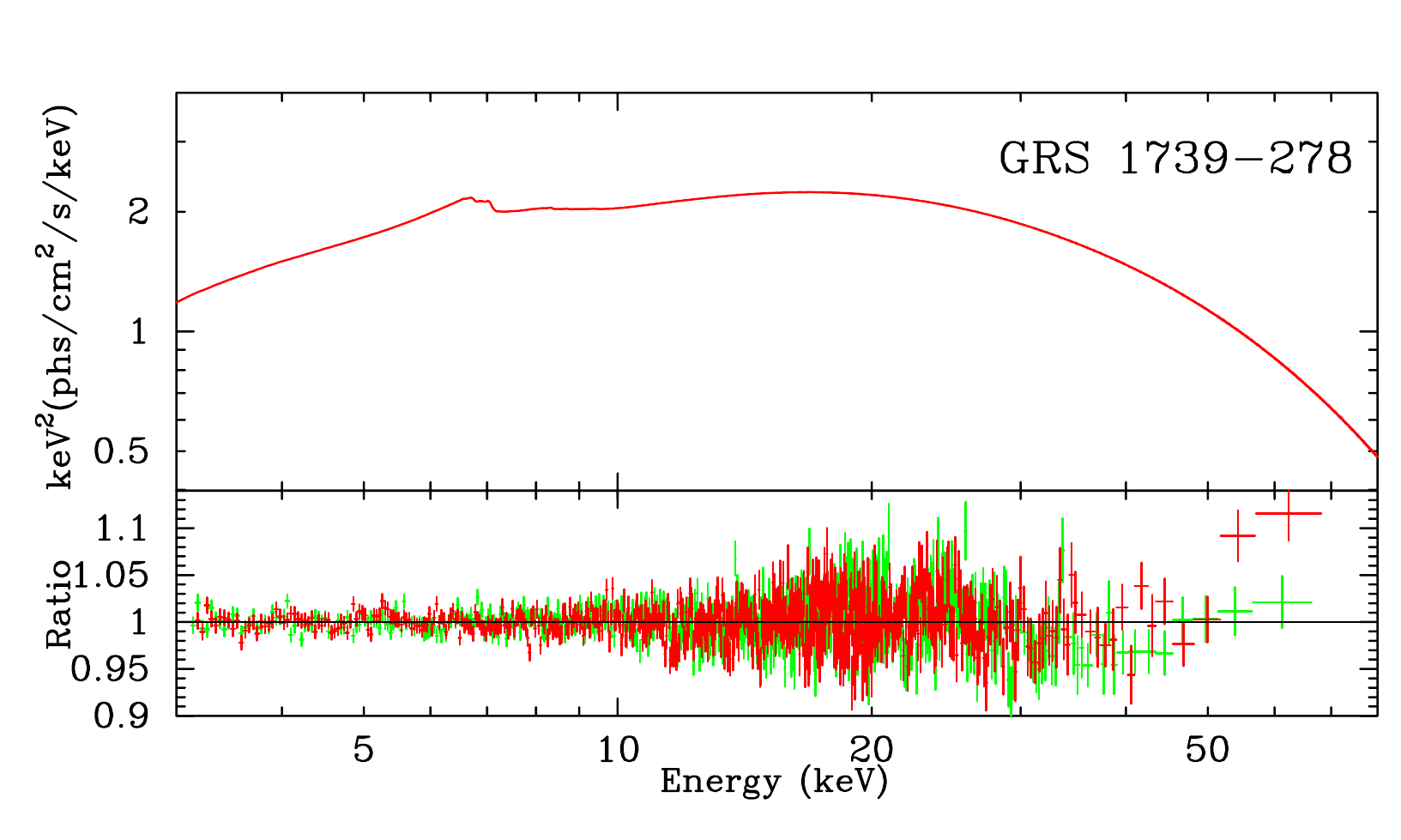}
\hspace{0.25cm}
\includegraphics[width=8.5cm,trim={2.0cm 0.5cm 3.0cm 17.5cm},clip]{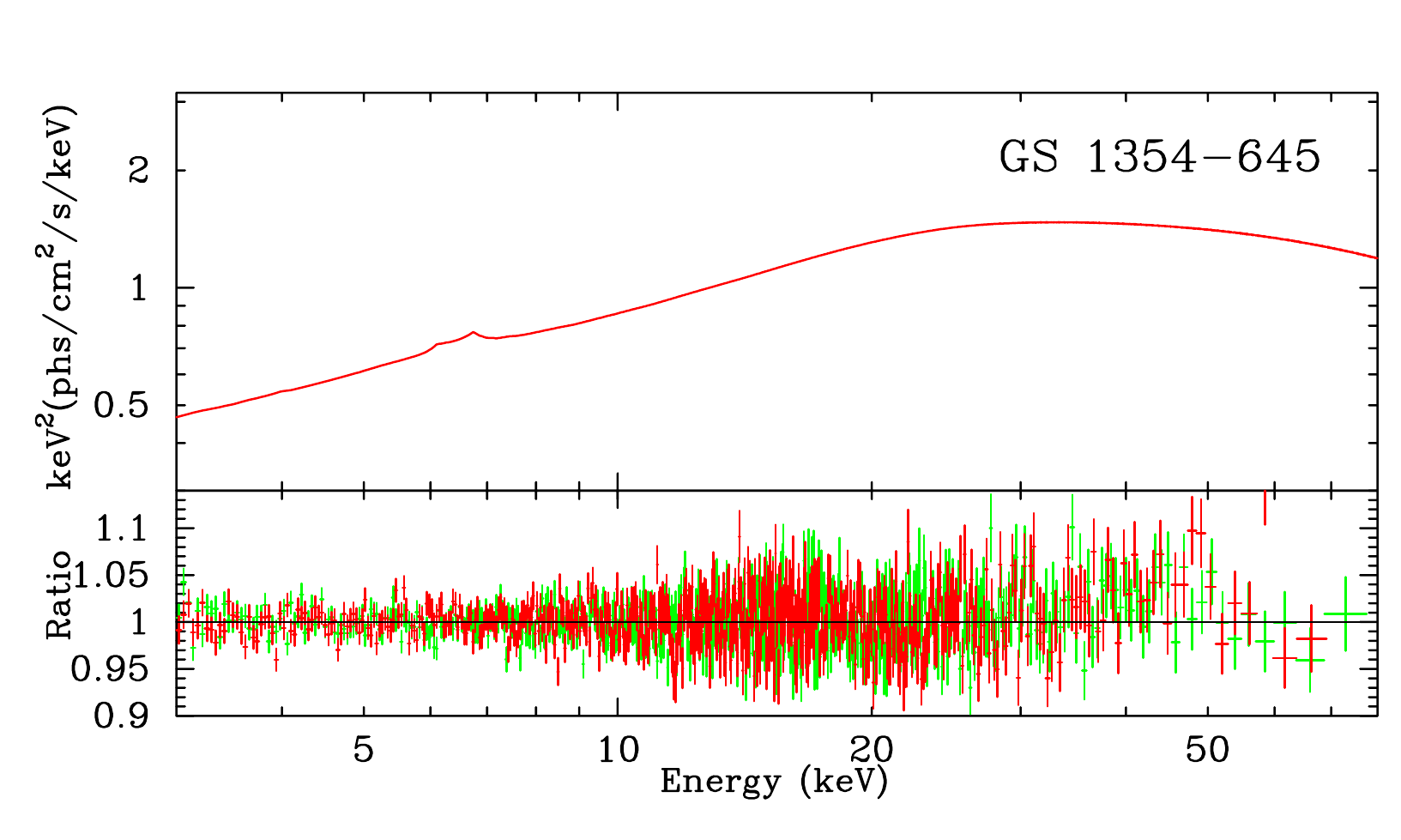} \\
\includegraphics[width=8.5cm,trim={2.0cm 0.5cm 3.0cm 17.5cm},clip]{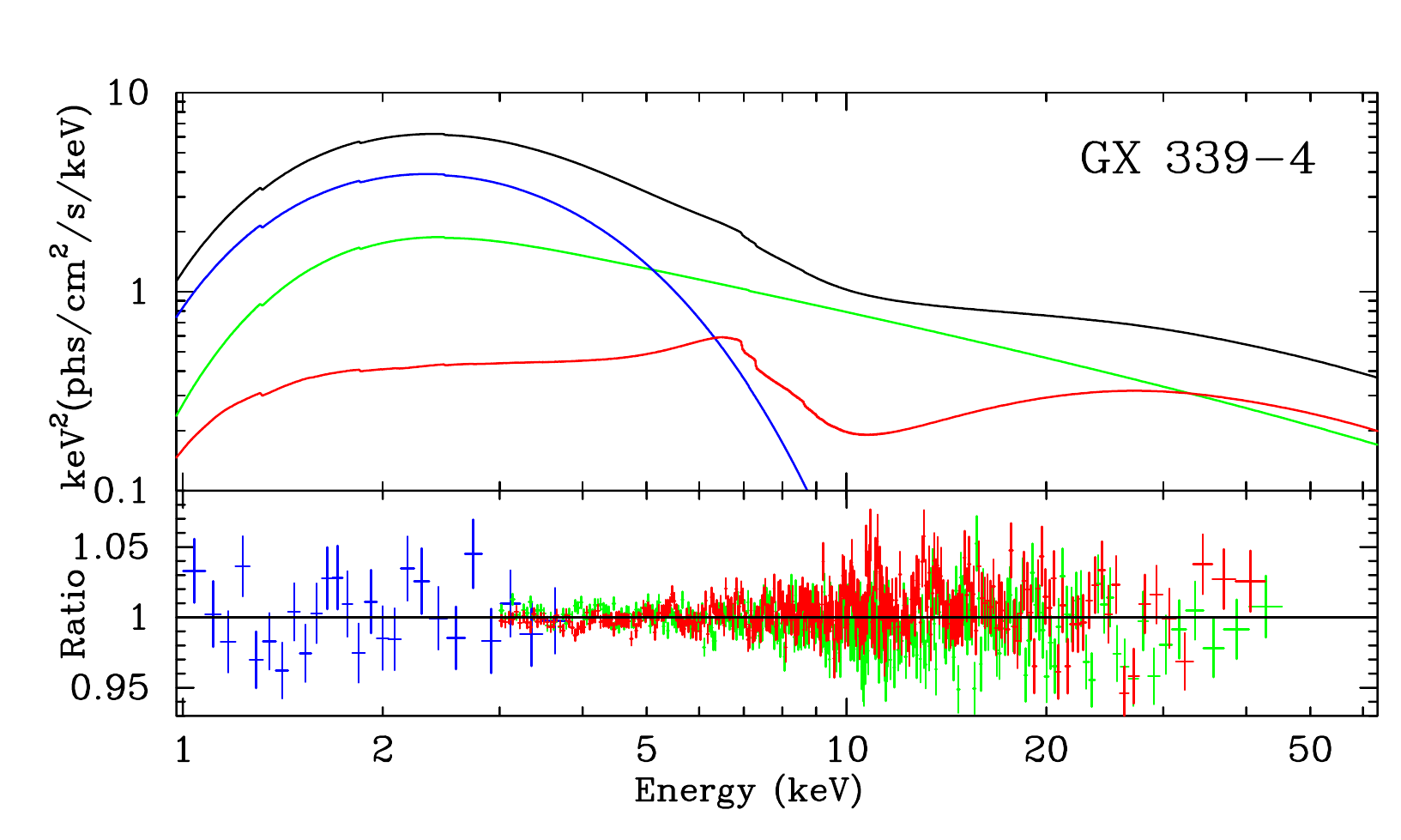}
\hspace{0.25cm}
\includegraphics[width=8.5cm,trim={2.0cm 0.5cm 3.0cm 17.0cm},clip]{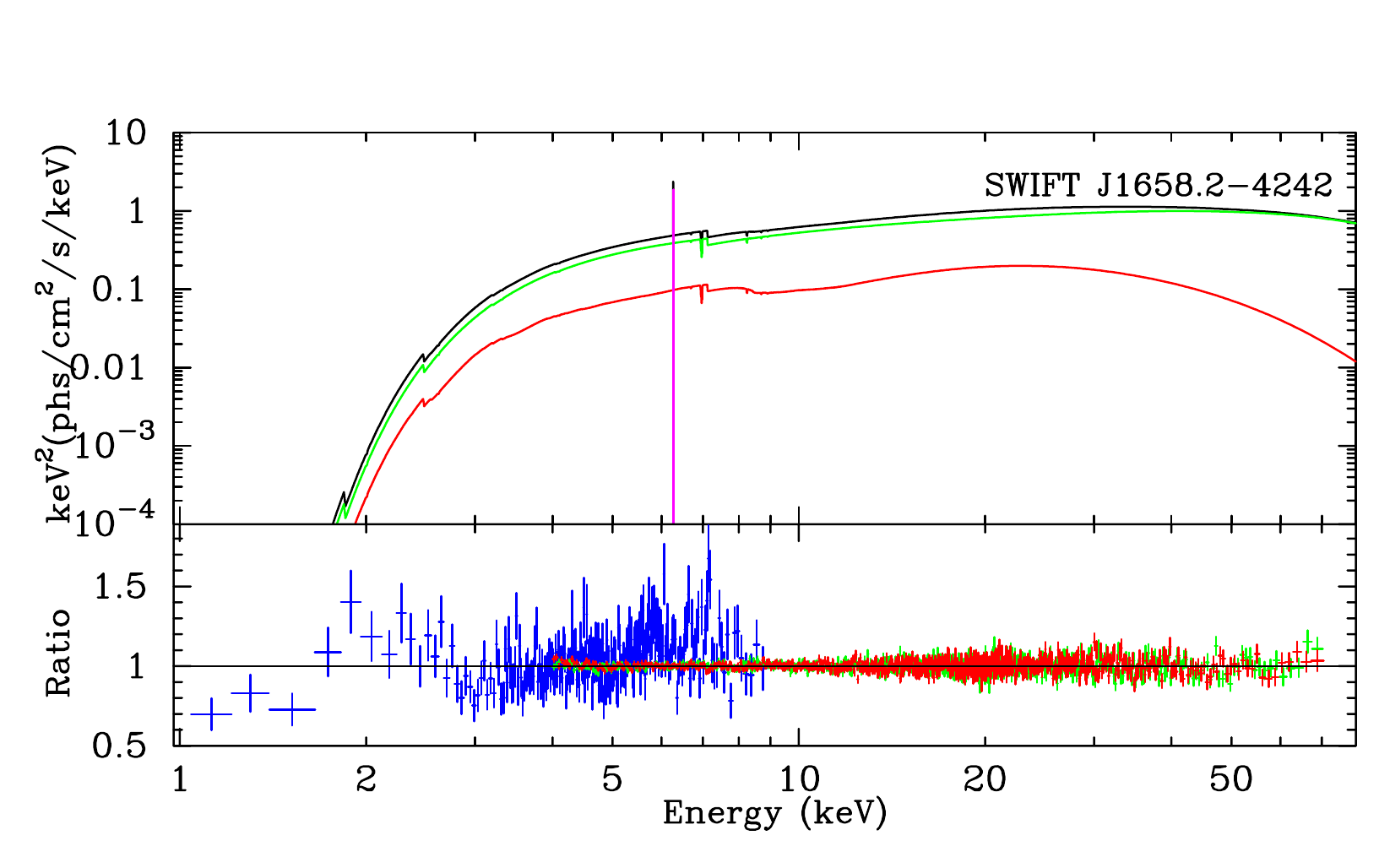}
\end{center}
\vspace{-0.5cm}
\caption{Best-fit model and data to best-fit model ratio for the spectra analyzed in this work. For the best-fit model plots, the black curve (when present) refers to the total model, the red curve to the relativistic reflection spectrum described by {\tt relxill(lp)(Cp)\_nk}, the orange curve (in EXO~1846--031) to the thermal component described by {\tt diskbb}, the blue curve (in GX~339--4) to the thermal model {\tt nkbb}, the green curve to the continuum {\tt comptt} (in GX~339--4) or {\tt nthComp} (in Swift~J1658.2--4242), and the magenta curve (in Swift~J1658.2--4242) is an emission line. In the ratio plots, green crosses are for FPMA/\textsl{NuSTAR} data, red crosses are for FPMB/\textsl{NuSTAR} data, and blue crosses are for XRT/\textsl{Swift} data. \label{f-mr}}
\end{figure*}

\begin{figure*}
\begin{center}
\includegraphics[width=17.0cm,trim={0cm 0cm 0cm 0cm},clip]{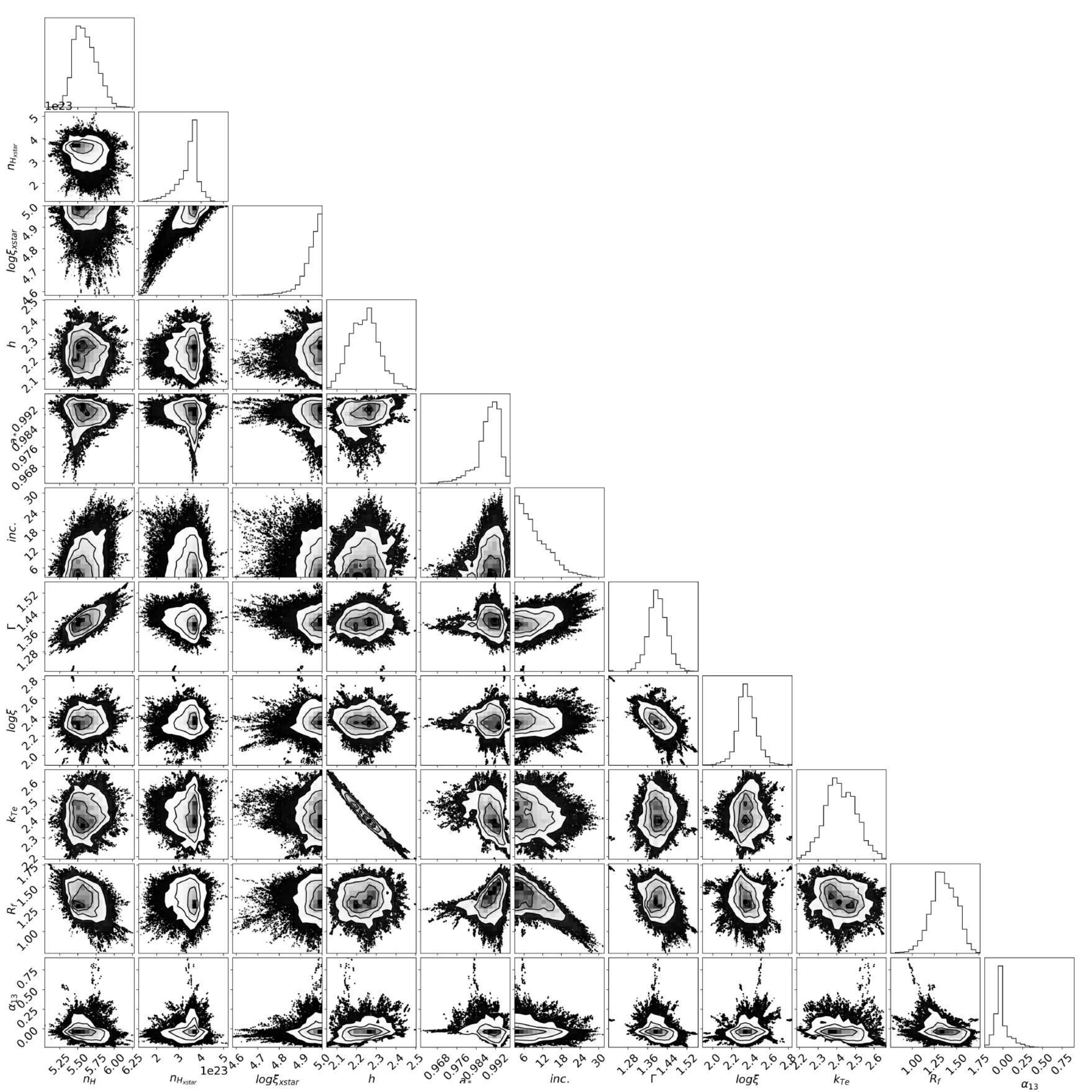}
\end{center}
\vspace{-0.5cm}
\caption{Corner plot for the free parameter-pairs (excluding calibration constants and component normalization) for 4U~1630--472 after the MCMC run. The 2D plots report the 1-, 2-, and 3-$\sigma$ confidence contours. \label{f-4u-mcmc}}
\end{figure*}

\begin{figure*}
\begin{center}
\includegraphics[width=17.0cm,trim={0cm 0cm 0cm 0cm},clip]{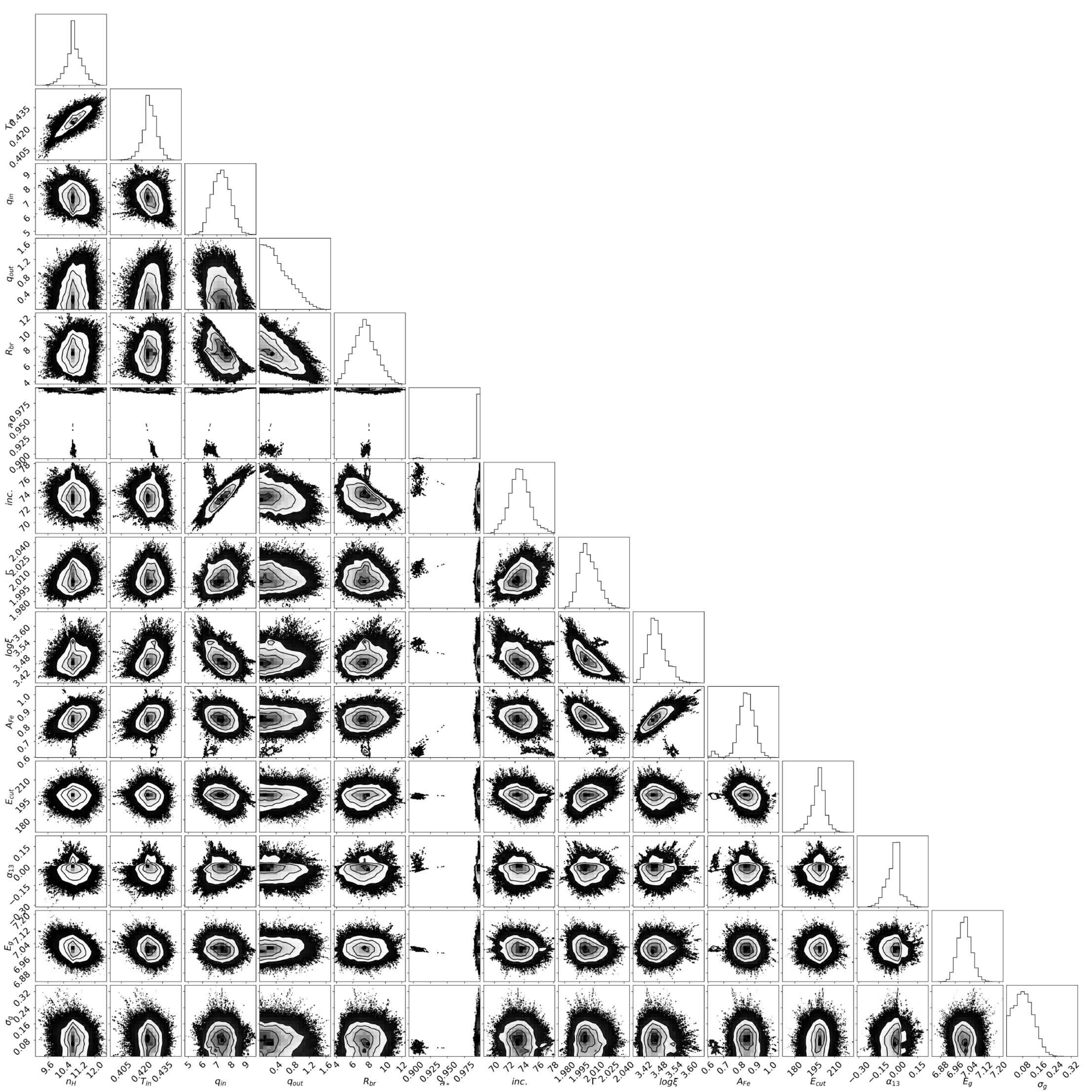}
\end{center}
\vspace{-0.5cm}
\caption{As in Fig.~\ref{f-4u-mcmc} for EXO~1846--031. \label{f-exo-mcmc}}
\end{figure*}

\begin{figure*}
\begin{center}
\includegraphics[width=17.0cm,trim={0cm 0cm 0cm 0cm},clip]{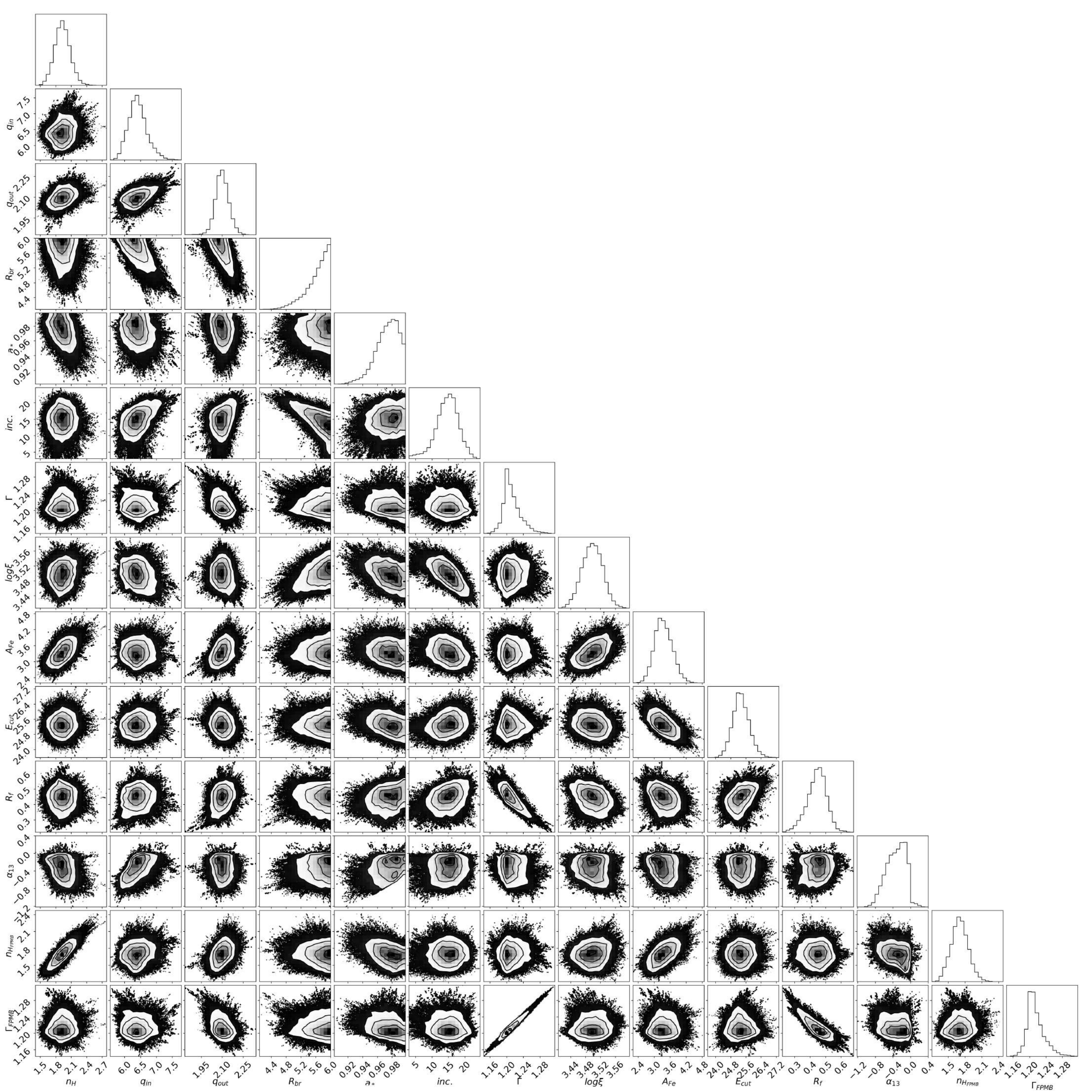}
\end{center}
\vspace{-0.5cm}
\caption{As in Fig.~\ref{f-4u-mcmc} for GRS~1739--278. \label{f-grs1739-mcmc}}
\end{figure*}

\begin{figure*}
\begin{center}
\includegraphics[width=17.0cm,trim={0cm 0cm 0cm 0cm},clip]{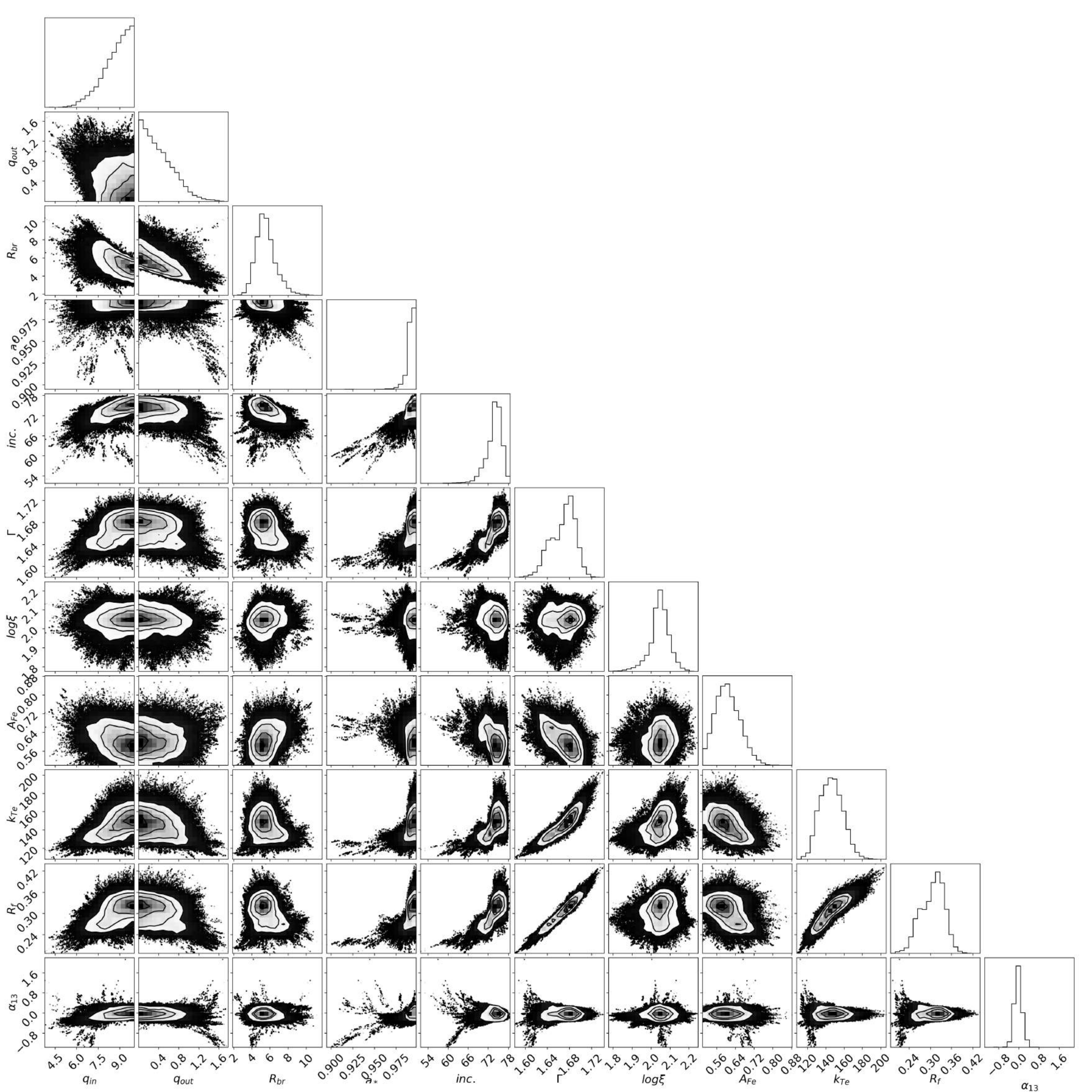}
\end{center}
\vspace{-0.5cm}
\caption{As in Fig.~\ref{f-4u-mcmc} for GS~1354--645. \label{f-gs1354-mcmc}}
\end{figure*}

\begin{figure*}
\begin{center}
\includegraphics[width=17.0cm,trim={0cm 0cm 0cm 0cm},clip]{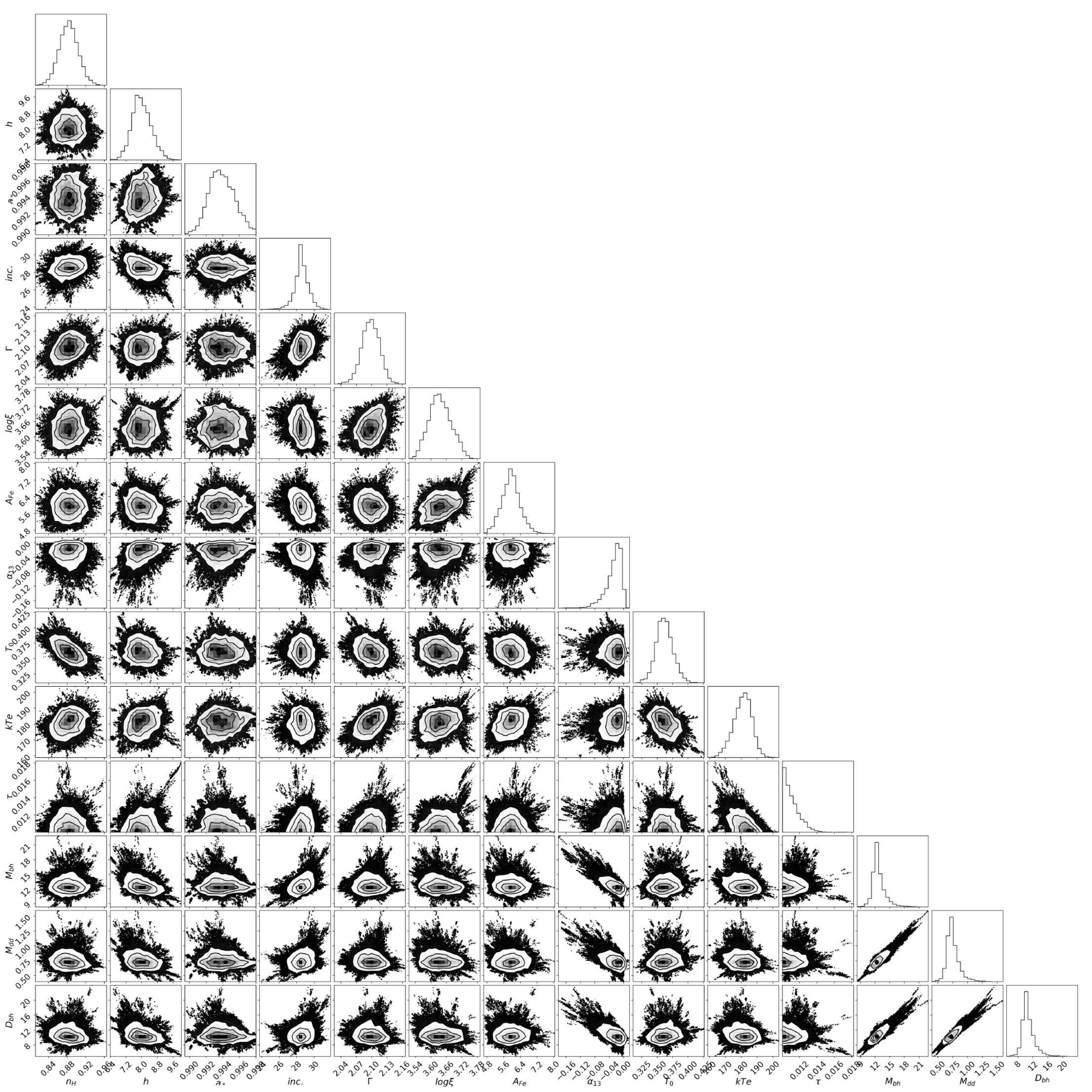}
\end{center}
\vspace{-0.5cm}
\caption{As in Fig.~\ref{f-4u-mcmc} for GX~339--4. \label{f-gx-mcmc}}
\end{figure*}

\begin{figure*}
\begin{center}
\includegraphics[width=17.0cm,trim={0cm 0cm 0cm 0cm},clip]{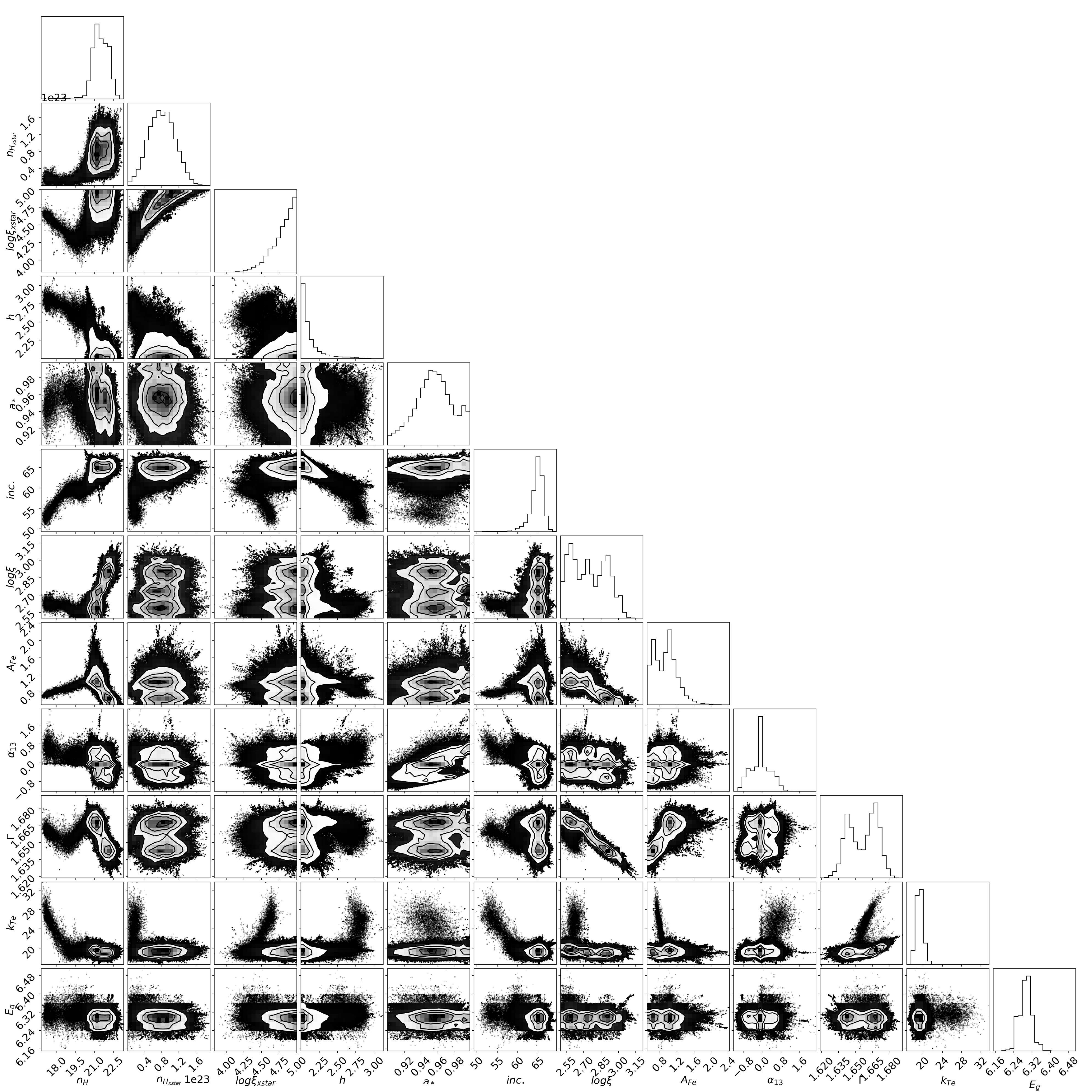}
\end{center}
\vspace{-0.5cm}
\caption{As in Fig.~\ref{f-4u-mcmc} for Swift~J1658.2--4242. \label{f-swift-mcmc}}
\end{figure*}

Since even for a single source we have several free parameters in the model, we run Markov Chain Monte-Carlo (MCMC) analyses. We use the python script by Jeremy Sanders employing {\tt emcee} (MCMC Ensemble sampler implementing Goodman \& Weare algorithm)\footnote{Available on github at\\ \url{https://github.com/jeremysanders/xspec\_emcee}.}. Our results are shown in Tab.~\ref{tab-fit}, where the reported uncertainties correspond to 1-$\sigma$ (for $\alpha_{13}$ we also show the 3-$\sigma$ estimate). The best-fit models and the data to best-fit model ratios of our six sources are shown in Fig.~\ref{f-mr}. The corner plots resulting from the MCMC analyses are shown in Figs.~\ref{f-4u-mcmc}-\ref{f-swift-mcmc}.

\begin{table*}
\centering
{\renewcommand{\arraystretch}{1.1}
\begin{tabular}{lcccccc}
\hline\hline
& 4U~1630--472 & EXO~1846--031 & GRS~1739--278 & GS~1354--645 &  GX~339--4 & Swift~J1658.2--4242 \\
\hline\hline
{\tt tbabs} \\
$N_{\rm H}$ [$10^{22}$~cm$^{-2}$] & $5.59^{+0.18}_{-0.13}$ & $10.9^{+0.4}_{-0.3}$ & $1.92^{+0.15}_{-0.14}$ & $0.7^*$ & $0.88^{+0.02}_{-0.02}$ & $21.4^{+0.7}_{-0.6}$ \\
& -- & -- & $1.70^{+0.15}_{-0.15}$ & -- & -- & -- \\
\hline 
{\tt xstar} \\
$N_{\rm H}$ [$10^{22}$~cm$^{-2}$] & $35.6^{+1.9}_{-5.9}$ & -- & -- & -- & -- & $79.3^{+0.3}_{-0.3}$ \\
$\log\xi$ [erg~cm~s$^{-1}$] & $4.96^{+0.03}_{-0.05}$ & -- & -- & -- & -- & $4.84^{+0.12}_{-0.21}$ \\
\hline
{\tt diskbb} \\
$kT_{\rm in}$ [keV] & -- & $0.427^{+0.007}_{-0.009}$ & -- & -- & -- & -- \\
\hline
{\tt nkbb} \\
$M$ [$M_\odot$] & -- & -- & -- & -- & $12.4^{+1.4}_{-0.7}$ & -- \\
$\dot{M}$ [$10^{18}$~g/s] & -- & -- & -- & -- & $0.73^{+0.11}_{-0.07}$ & -- \\
$D$ [kpc] & -- & -- & -- & -- & $10.4^{+1.8}_{-1.0}$ & -- \\
\hline
{\tt comptt} && \\
$kT_0$ [keV] &  -- & -- & -- & -- & $0.355^{+0.014}_{-0.011}$ & -- \\
$kT_{\rm plasma}$ [keV] &  -- & -- & -- & -- & $181^{+5}_{-6}$ & -- \\
$\tau$ & -- & -- & -- & -- & $<0.01$ & -- \\
\hline
{\tt nthComp} && \\
$kT_0$ [keV] &  -- & -- & -- & -- & -- & $0.1^*$ \\
$kT_{\rm plasma}$ [keV] &  -- & -- & -- & -- & -- & $38.6^{+0.7}_{-0.7}$ \\
\hline
{\tt relxill(lp)(Cp)\_nk} \\
$h$ $[r_{\rm g}]$ & $2.24^{+0.07}_{-0.08}$ & -- & -- & -- & $8.0^{+0.5}_{-0.4}$ & $2.07^{+0.17}_{-0.05}$ \\
$q_{\rm in}$ & -- & $7.3^{+0.6}_{-0.6}$ & $6.4^{+0.3}_{-0.3}$ & $8.9^{+0.8}_{-1.2}$ & -- & -- \\
$q_{\rm out}$ & -- & $0.4^{+0.4}_{-0.3}$ & $2.10^{+0.05}_{-0.04}$ & $0.4^{+0.4}_{-0.3}$ & -- & -- \\
$R_{\rm br}$ [$r_{\rm g}$] & -- & $7.4^{+1.5}_{-1.4}$ & $5.70^{+0.21}_{-0.39}$ & $5.4^{+1.2}_{-0.9}$ & -- & -- \\
$a_*$ & $0.990^{+0.003}_{-0.005}$ & $0.9968^{+0.0015}_{-0.0017}$ & $0.975^{+0.014}_{-0.018}$ & $0.992^{+0.004}_{-0.004}$ & $0.9944^{+0.0009}_{-0.0015}$ & $0.955^{+0.024}_{-0.022}$ \\
$i$ [deg] & $8^{+6}_{-4}$ & $73.4^{+1.3}_{-1.3}$ & $15^{+3}_{-3}$ & $74.1^{+1.7}_{-2.5}$ & $28.5^{+0.78}_{-0.7}$ & $64.9^{+1.1}_{-1.6}$ \\
$\Gamma$ & $1.40^{+0.04}_{-0.03}$ & $2.002^{+0.018}_{-0.013}$ & $1.21^{+0.02}_{-0.02}$ & $1.67^{+0.02}_{-0.03}$ & $2.10^{+0.03}_{-0.03}$ & $1.667^{+0.008}_{-0.023}$ \\
$E_{\rm cut}$ [keV] & -- & $198^{+4}_{-5}$ & $25.3^{+0.5}_{-0.4}$ & -- & $2 g k T_{\rm plasma}$ & -- \\
$k T_{\rm e}$ [keV] & $2.42^{+0.09}_{-0.08}$ & -- & -- & $146^{+13}_{-13}$ & -- & $g k T_{\rm plasma}$ \\ 
$\log\xi$ [erg~cm~s$^{-1}$] & $2.36^{+0.10}_{-0.08}$ & $3.47^{+0.04}_{-0.03}$ & $3.50^{+0.03}_{-0.03}$ & $2.05^{+0.04}_{-0.04}$ & $3.63^{+0.05}_{-0.04}$ & $2.73^{+0.18}_{-0.16}$ \\
$A_{\rm Fe}$ & $1^*$ & $0.85^{+0.05}_{-0.05}$ & $3.3^{+0.4}_{-0.3}$ & $0.60^{+0.06}_{-0.05}$ & $5.9^{+0.4}_{-0.4}$ & $0.95^{+0.23}_{-0.31}$ \\
$R_{\rm f}$ & $1.35^{+0.15}_{-0.14}$ & $1^*$ & $0.45^{+0.05}_{-0.07}$ & $0.31^{+0.03}_{-0.04}$ & -- & -- \\
$\alpha_{13}$ & $-0.03^{+0.08}_{-0.06}$ & $-0.03^{+0.03}_{-0.06}$ & $-0.31^{+0.20}_{-0.25}$ & $-0.04^{+0.16}_{-0.15}$ & $-0.023^{+0.015}_{-0.026}$ & $-0.05^{+0.44}_{-0.47}$ \\
$\alpha_{13}$ (3$\sigma$) & $-0.03^{+0.63}_{-0.18}$ & $-0.03^{+0.17}_{-0.18}$ & $-0.31^{+0.64}_{-0.53}$ & $-0.04^{+0.58}_{-0.87}$ & $-0.023^{+0.030}_{-0.137}$ & $-0.05^{+1.21}_{-0.97}$ \\
\hline
{\tt gaussian} \\
$E_{\rm line}$ & -- & $7.01^{+0.04}_{-0.04}$ & -- & -- & -- & $6.29^{+0.02}_{-0.03}$ \\
\hline
$\chi^2_\nu$ & 1224.01/1084 & 2751.90/2594 & 1285.84/1112 & 2895.14/2720 & 869.15/647 & 1741.78/1644 \\
& =1.1292 & =1.0609 & =1.1563 & =1.0644 & =1.3434 & =1.0595 \\ 
\hline\hline
\end{tabular}
}
\caption{\rm Summary of the best-fit values of the six sources analyzed in this work. \label{tab-fit}}
\end{table*}

\begin{figure}[t]
\begin{center}
\includegraphics[width=8.5cm,trim={2.5cm 0.5cm 3.0cm 14.5cm},clip]{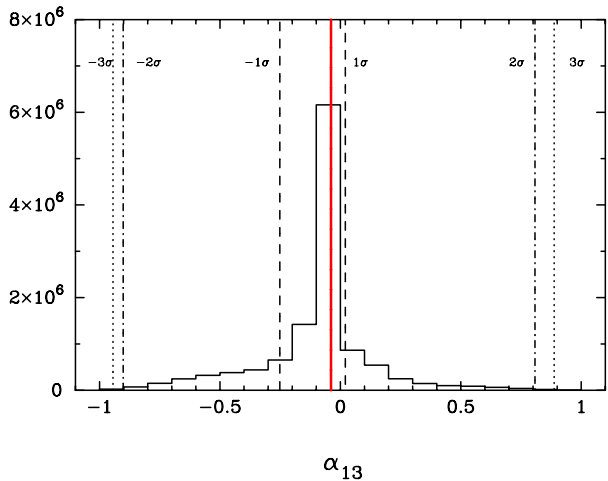}
\end{center}
\vspace{-0.6cm}
\caption{Histogram of the Johannsen deformation parameter $\alpha_{13}$ resulting from the combination of the measurements from the six sources analyzed in this work. The red solid line marks the estimate of $\alpha_{13}$ and the dashed, dashed-dotted, and dotted lines mark, respectively, the 1-, 2-, and 3-$\sigma$ limits. \label{f-a13}}
\end{figure}

\begin{table*}
\centering
{\renewcommand{\arraystretch}{1.1}
\begin{tabular}{lcccccc}
\hline\hline
& 4U~1630--472 & EXO~1846--031 & GRS~1739--278 & GS~1354--645 &  GX~339--4 & Swift~J1658.2--4242 \\
\hline\hline
{\tt xillver} \\
$\chi^2_\nu$ & 2609.14/1104 & 2856.58/2599 & 1453.92/1118 & 3116.80/2725 & 1521.35/649 & 1858.85/1650 \\
& =2.363 & =1.099 & =1.300 & =1.144 & =2.344 & =1.126\\
\hline 
{\tt relxill(Cp)\_nk} \\
$\alpha_{13}$ & $0.16_{-0.31}^{+0.31}$ & $-0.03_{-0.18}^{+0.17}$ & $-0.31_{-0.53}^{+0.64}$ & $-0.04_{-0.87}^{+0.58}$ & $-0.010_{-0.104}^{+0.066}$ &$-0.03_{-0.39}^{+0.19}$ \\
$\chi^2_\nu$ & 1273.74/1082 & 2751.90/2594 & 1283.78/1112 & 2895.07/2720 & 874.22/645 & 1754.14/1642 \\
& =1.177 & =1.061 & =1.154 & =1.065 & =1.355 & =1.068 \\
\hline 
{\tt relxilllp(Cp)\_nk} \\
$\alpha_{13}$ & $-0.03_{-0.18}^{+0.63}$ & $2.7_{-3.9}^{+1.8}$ & $0.4_{-4.9}^{+4.1}$& $0.4_{-1.9}^{+2.1}$ & $-0.023_{-0.030}^{+0.137}$ & $-0.05_{-0.97}^{+1.21}$ \\
$\chi^2_\nu$ & 1224.01/1084 & 2773.72/2596 & 1326.45/1114 & 2936.96/2722 & 869.15/647 & 1741.81/1644 \\
& =1.129 & =1.068 & =1.190 & =1.079 & =1.346 & =1.059 \\
\hline\hline
\end{tabular}
}
\caption{\rm Summary of $\chi^2$s and 3-$\sigma$ measurements of $\alpha_{13}$ modeling the reflection component with {\tt xillver}, {\tt relxill(Cp)\_nk}, and {\tt relxilllp(Cp)\_nk}. \label{tab-fit2}}
\vspace{0.5cm}
\end{table*}

\subsubsection{Combined measurement}

If we assume that the value of $\alpha_{13}$ is the same for all black holes, we can combine the six individual measurements to get a joint measurement. Such a possibility would depend on the (unknown) underlying theory. Within the phenomenological framework of the Johannsen metric, it is just an assumption. If astrophysical black holes had more ``hairs'' than those predicted by General Relativity (e.g. some extra charge), deviations from the Kerr background would be different for every object. In such a case, we should not combine the data to get a joint measurement of $\alpha_{13}$. If some form of the no-hair theorems would still hold in Nature, all astrophysical black holes may share the same deviation from the Kerr solution and the deformation parameter may have the same value for all sources (or very similar value for all sources of the same class, like stellar-mass black holes). In this section, we thus assume this second scenario.

There are also different methods to combine the individual measurements to get a joint measurement of the deformation parameter $\alpha_{13}$. Here we combine the distribution of the six measurements and we get the histogram in Fig.~\ref{f-a13}. The 1-$\sigma$ joint measurement is
\be
\alpha_{13} = -0.04_{-0.21}^{+0.06} \, .
\ee 
The 3-$\sigma$ measurement is $\alpha_{13} = -0.04_{-0.90}^{+0.93}$.


\section{Discussion and conclusions} \label{s:dis}

\begin{figure}[b]
\begin{center}
\includegraphics[width=8.5cm,trim={0.6cm 0cm 0.4cm 0cm},clip]{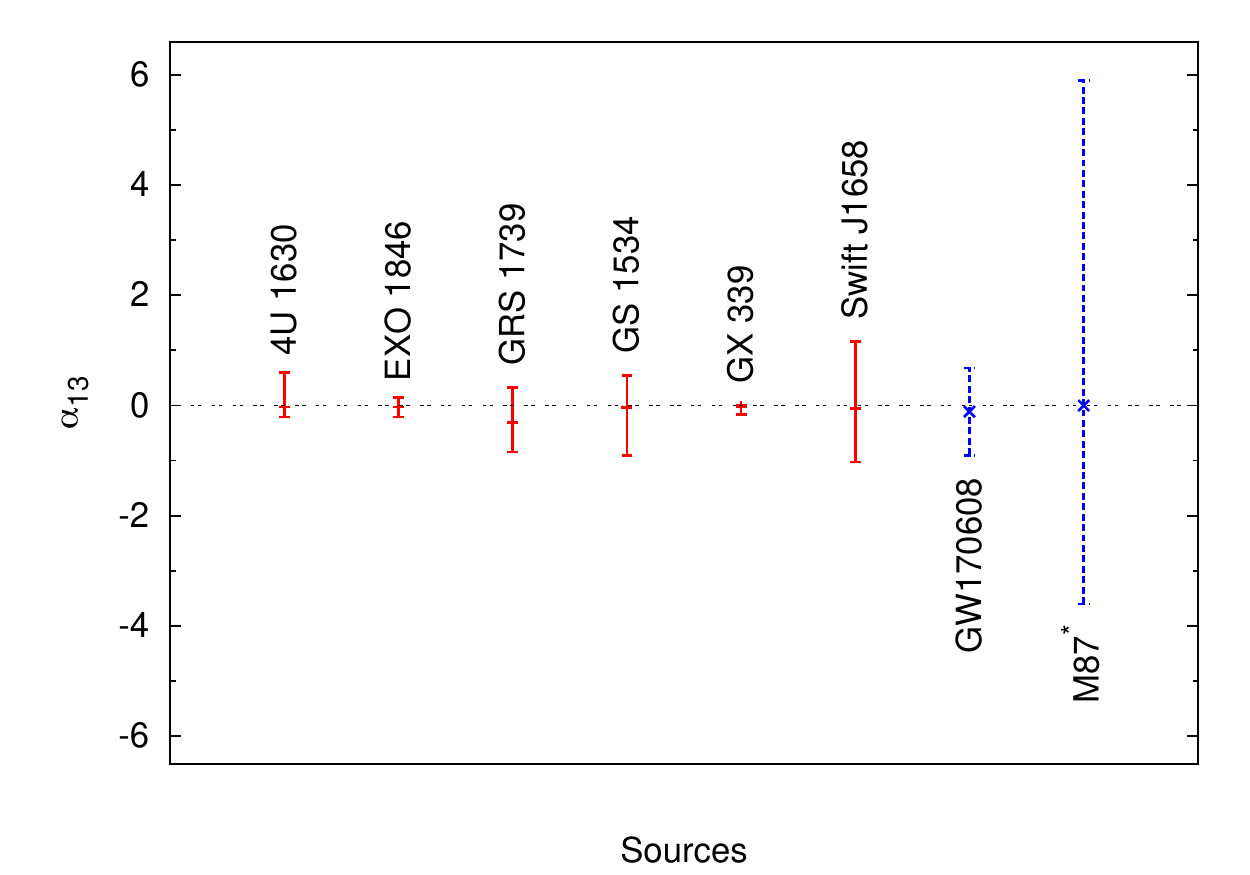}
\end{center}
\vspace{-0.5cm}
\caption{Summary of the 3-$\sigma$ measurements on the Johannsen deformation parameter $\alpha_{13}$ from the six sources analyzed in this work. We also show the 3-$\sigma$ constraint from GW170608~\citep{Cardenas-Avendano:2019zxd} and the 1-$\sigma$ constraint from M87$^*$~\citep{Psaltis:2020lvx}. \label{f-summary}}
\end{figure}

In our study, we employed state-of-the-art in relativistic reflection modeling to analyze the currently available best X-ray data of reflection dominated spectra of black hole binaries to obtain precise and accurate constraints on the Kerr black hole hypothesis. Our results turn out to be in perfect agreement with the predictions of General Relativity and are summarized in Fig.~\ref{f-summary}, where we show the 3-$\sigma$ measurements of $\alpha_{13}$ for every source.

Our constraints are much stronger than that recently obtained from the 2017 observations of the Event Horizon Telescope of the supermassive black holes in the galaxy M87, which is $-3.6 < \alpha_{13} < 5.9$ at 1-$\sigma$~\citep{Psaltis:2020lvx}. Employing the Einstein Equations for the production of gravitational waves, it is possible to constrain $\alpha_{13}$ from the LIGO/Virgo data~\citep{Cardenas-Avendano:2019zxd}. The strongest constraint was obtained from GW170608 and the 3-$\sigma$ measurement is $\alpha_{13} = -0.11 \pm 0.79$, which is shown together with our constraints from \textsl{NuSTAR} data in Fig.~\ref{f-summary}. The constraints obtained here are at worst comparable with the strongest constraint obtained from gravitational waves. For some sources (GX~339--4 and EXO~1846--031), our constraints are much stronger\footnote{We remind the reader that, within a phenomenological framework, like that employed in the present work, a comparison between electromagnetic and gravitational wave constraints should be taken with caution. Gravitational wave tests can often get stringent constraints from the dynamical aspects of the theory (dipolar radiation, ringdown, inspiral-merger, consistency tests...), which are not possible with only the assumption of a background metric.}.

We note that the constraints obtained with our technique strongly depend on the specific source and observation, because the morphology of the accretion material and the spectrum of the source can significantly change from source to source and even for the same source from an outburst to another outburst. Gravitational wave tests are less sensitive to the specific event because, in any case, every event is the coalescence of two black holes and effects of the surrounding environment are normally negligible.

To clarify the significance of a relativistic reflection component (again a non-relativistic reflection component that could be modeled by {\tt xillver}), as well as the impact of the emissivity profile (i.e., broken power-law vs lamppost geometry), in Tab.~\ref{tab-fit2} we show $\chi^2$ and measurement of $\alpha_{13}$ for different fits. For all spectra, {\tt xillver} does not provide an acceptable fit. When we employ a relativistic reflection component, we choose the model providing the lowest $\chi^2$ as the best model. We note that, when the best model is {\tt relxilllp(Cp)\_nk} (lamppost geometry), the model with a broken power-law does not provide a too bad fit and the measurement of $\alpha_{13}$ is consistent with that obtained with the lamppost model. The contrary is not true: if the best model is {\tt relxill(Cp)\_nk} (broken power-law emissivity profile), the model with a lamppost geometry cannot constrain the deformation parameter well. Such a result is perfectly understandable: the broken power-law can approximate well the lamppost profile, while the latter may not approximate well the broken power-law. For the sources fit with a broken power-law (EXO~1846--031, GRS~1739--278, and GS~1354--645), we find $q_{\rm out}$ quite low, while at large radii the profile of the lamppost model is similar to a power-law emissivity profile with emissivity index close to 3.

The models employed in our study to fit the spectra of the six sources present a number of simplifications that could, potentially, introduce undesirable systematic uncertainties to make our constraints on the Johannsen deformation parameter $\alpha_{13}$ unreliable. Here we point out that the main systematic uncertainties are well under control for these specific sources.

\begin{enumerate}

\item {\it Inner edge of the accretion disk}. In our measurements of the Johannsen deformation parameter $\alpha_{13}$ reported in this work, we assume that the inner edge of the accretion disk is at the ISCO radius. While there is a common consensus that this is a good approximation for sources in the soft state with an Eddington-scaled disk luminosity between $\sim$5\% to $\sim$30\%~\citep{Steiner:2010kd,Kulkarni:2011cy}, the issue is more controversial for sources in the hard state and the disk may be truncated at a larger radius. We note, however, that this assumption has a negligible impact on our measurements of $\alpha_{13}$ because, assuming $R_{\rm in} = R_{\rm ISCO}$, we find that all our sources are very-fast-rotating black holes, with a value of the spin parameter $a_*$ close to the maximum limit. The inner edges of the accretion disks of these sources is very close to the black hole. Fitting the data with $R_{\rm in}$ as a free parameter, we recover the same results.  

\item {\it Thickness of the disk}. The versions of {\tt relxill\_nk} and {\tt nkbb} employed in our analysis assume that the accretion disk is infinitesimally thin, while in reality the disk has a finite thickness, which increases as the mass accretion rate increases. Reflection and thermal models with accretion disks of finite thickness were presented in \citet{Taylor:2018rlv,Abdikamalov:2020oci,Zhou:2020koa} and their impact on the estimate of the model parameters was explored. Again, since we found that our sources have a spin parameter close to the maximum limit, the radiative efficiency is very high and the actual thickness of the disk is modest. In such a case, the impact on the constraints on $\alpha_{13}$ is negligible, as it was explicitly shown in~\citet{Abdikamalov:2020oci}.

\item {\it Radiation from the plunging region}. In our model to fit the data, we ignore the radiation emitted from the plunging region. If the Eddington-scaled mass accretion rate is not very low, below 1\% for fast-rotating black holes, the plunging region is optically thick~\citep{Reynolds:1997ek}. In such a case, the Comptonized photons from the corona should generate a reflection component. However, the gas is highly ionized in the plunging region~\citep{Reynolds:1997ek,Wilkins:2020pgu}. The impact of such radiation on the estimate of the Johannsen deformation parameter $\alpha_{13}$ was studied in \citet{Cardenas-Avendano:2020xtw}, where it was shown that for fast-rotating black holes associated with very small plunging regions (which is the case of the sources presented in our study) the impact is negligible for present and near future X-ray data. 

\item {\it Disk emissivity profile}. The emissivity profile of the accretion disk is an important ingredient to properly model the reflection spectrum of the source and an incorrect emissivity profile can lead to a non-vanishing deformation parameter $\alpha_{13}$~\citep{Zhang:2019ldz}. In our study, we tried power law, broken power law, and lamppost model, and eventually we selected the emissivity profile providing the best fit. As we have already discussed, when the best model is the lamppost geometry, the measurement with the broken power-law model is not too different, as the the latter is a more flexible profile and can approximate the lamppost one well. The contrary is not true, and if the spectrum prefers a broken power-law, the fit and the constraints with the lamppost model may be bad.

\item {\it Disk electron density}. We fit the data assuming a disk electron density $n_{\rm e} = 10^{15}$~cm$^{-3}$, which is the default value in {\tt relxil\_nk}. However, such a value is likely too low for the accretion disk of stellar-mass black holes with a relatively high mass accretion rate, which is the case of our sources~\citep{Jiang:2019xqn,Jiang:2019ztr}. The impact of the disk electron density on the estimate of $\alpha_{13}$ was explored in previous studies~\citep{Tripathi:2020dni,Zhang:2019ldz}, where we did not find any clear bias in the constraint on $\alpha_{13}$ from the value of $n_{\rm e}$, which is consistent with the conclusion in~\citet{Jiang:2019xqn,Jiang:2019ztr} that underestimating the disk electron density has no significant impact on the estimate of the black hole spin (assuming the Kerr metric).  

\item {\it Returning radiation}. Because of the strong light bending near the black hole, reflection radiation emitted from the accretion disk may return to the accretion disk and be reflected again. Since the reflection spectrum depends on the incident one, the reflection radiation returning to the disk cannot be reabsorbed in the emissivity profile induced by the corona. At the moment, there are only some preliminary and incomplete studies on the impact of the returning radiation on the estimate of the model parameters when the effect is not included in the calculations. The effect is more important for sources with the inner edge of the accretion disk very close to the black hole, which is the case for our sources. While we admit that the systematic uncertainties of such an effect is not well under control as of now, we note that it is important for low inclination angles and quite small for edge-on disks~\citep{Riaz:2020zqb}. In other words, we do not expect that the returning radiation has any relevant impact on our analysis of EXO~1846--031, GS~1354--645, and Swift~J1658--4242, while we are currently unable to quantify the impact on the constraint on $\alpha_{13}$ for 4U~1630--472, GRS~1739--278, and GX~339--4. 

\end{enumerate}

Last, we wish to point out that our constraints can be improved with the next generation of X-ray observatories. \textsl{eXTP}~\citep{Zhang:2016ach} will provide simultaneous timing, spectral, and polarization measurements of the disk reflection component. \textsl{Athena}~\citep{Nandra:2013jka} will have a much better energy resolution near the iron line.


\vspace{0.5cm}

{\bf Acknowledgments --}
We thank Alejandro Cardenas-Avendano for useful discussions about the constraints from the LIGO/Virgo data.
This work was supported by the Innovation Program of the Shanghai Municipal Education Commission, Grant No.~2019-01-07-00-07-E00035, the National Natural Science Foundation of China (NSFC), Grant No.~11973019, and Fudan University, Grant No.~JIH1512604. 
Y.Z. acknowledges the support from China Scholarship Council (CSC 201906100030).
D.A. is supported through the Teach@T{\"u}bingen Fellowship.
J.J. acknowledges the support from Tsinghua Shuimu Scholar Program and Tsinghua Astrophysics Outstanding Fellowship.
A.T., C.B., J.J., and H.L. are members of the International Team~458 at the International Space Science Institute (ISSI), Bern, Switzerland, and acknowledge support from ISSI during the meetings in Bern.


\appendix

\section{A. Johannsen metric}\label{a:johannsen}

For the convenience of the readers, we report here the expression of the Johannsen metric, but more details can be found in the original paper~\citep{Johannsen:2015pca}. In Boyer-Lindquist-like coordinates, the line element is
\be\label{eq-jm}
ds^2 &=&-\frac{\tilde{\Sigma}\left(\Delta-a^2A_2^2\sin^2\theta\right)}{B^2}dt^2 
+\frac{\tilde{\Sigma}}{\Delta A_5}dr^2+\tilde{\Sigma} d\theta^2 \nonumber\\
&& -\frac{2a\left[\left(r^2+a^2\right)A_1A_2-\Delta\right]\tilde{\Sigma}\sin^2\theta}{B^2}dtd\phi 
+\frac{\left[\left(r^2+a^2\right)^2A_1^2-a^2\Delta\sin^2\theta\right]\tilde{\Sigma}\sin^2\theta}{B^2}d\phi^2 \, ,
\ee
where $M$ is the black hole mass, $a = J/M$, $J$ is the black hole spin angular momentum, $\tilde{\Sigma} = \Sigma + f$, and
\be
\Sigma = r^2 + a^2 \cos^2\theta \, , \qquad
\Delta = r^2 - 2 M r + a^2 \, , \qquad
B = \left(r^2+a^2\right)A_1-a^2A_2\sin^2\theta \, .
\ee
The functions $f$, $A_1$, $A_2$, and $A_5$ are defined as
\be\label{eq-fa1a2a5}
f = \sum^\infty_{n=3} \epsilon_n \frac{M^n}{r^{n-2}} \, , \quad
A_1 = 1 + \sum^\infty_{n=3} \alpha_{1n} \left(\frac{M}{r}\right)^n \, , \quad
A_2 = 1 + \sum^\infty_{n=2} \alpha_{2n}\left(\frac{M}{r}\right)^n \, , \quad
A_5 = 1 + \sum^\infty_{n=2} \alpha_{5n}\left(\frac{M}{r}\right)^n \, ,
\ee
where $\{ \epsilon_n \}$, $\{ \alpha_{1n} \}$, $\{ \alpha_{2n} \}$, and $\{ \alpha_{5n} \}$ are four infinite sets of deformation parameters without constraints from the Newtonian limit and Solar System experiments. In the present study, we have only considered the deformation parameter $\alpha_{13}$, which has the strongest impact on the reflection spectrum. All other deformation parameters are set to zero.

We remind the readers that the Johannsen metric is a phenomenological deformation from the Kerr solution, and specifically designed for testing the Kerr black hole hypothesis with electromagnetic data. There is no theory from which the metric is derived, so no {\it a priori} indications about the value of these deformation parameters. The metric is obtained by imposing that the spacetime is regular outside of the event horizon (no naked singularities, closed time-like curves, etc.) and has a Carter-like constant (i.e., the equations of motion of a test-particle can be written in first-order form). Under these conditions, we get a metric characterized by four free functions ($f$, $A_1$, $A_2$, and $A_5$), which are then written in the form shown in Eq.~(\ref{eq-fa1a2a5}). The impact of the deformation parameters $\epsilon_3$, $\alpha_{13}$, $\alpha_{22}$, and $\alpha_{52}$ on a narrow line was shown in \citet{Bambi:2016sac}. Higher order deformation parameters (i.e. $\epsilon_n$ with $n \ge 4$, $\alpha_{1n}$ with $n \ge 4$, etc.) have a qualitatively very similar impact on a narrow line and on the full reflection spectrum. Our choice of considering only the deformation parameter $\alpha_{13}$ can thus be seen as a proof of concept of our current capabilities and our results should not be generalized to any generic deformation from the Kerr background.

\section{B. The {\tt relxill\_nk} package}\label{a:relxillnk}

{\tt relxill\_nk}~\citep{Bambi:2016sac,Abdikamalov:2019yrr} is an extension of the {\tt relxill} package~\citep{Dauser:2013xv,Garcia:2013lxa} to non-Kerr spacetimes. The default flavor is {\tt relxill\_nk}, but in the spectral analysis of our work we have also used {\tt relxilllp\_nk}, {\tt relxillCp\_nk}, and {\tt relxilllpCp\_nk}.

The spacetime metric is described by the black hole spin parameter $a_*$ and by the Johannsen deformation parameter $\alpha_{13}$, while the black hole mass does not directly enter the calculations of the reflection spectrum. The position of the observer is specified by the inclination angle of the disk with respect to the line of sight of the observer, $i$. In the default flavor {\tt relxill\_nk}, there are no assumptions about the coronal geometry and thus the emissivity profile of the disk is modeled with a broken power-law with three parameters: inner emissivity index $q_{\rm in}$, outer emissivity index $q_{\rm out}$, and breaking radius $R_{\rm br}$. In the lamppost version of the model (the flavors with {\tt lp}), we assume a lamppost coronal geometry: the corona is a point-like and isotropic source along the black hole spin axis and we have only one parameter, the coronal height $h$. In all flavors, the material of the accretion disk is characterized by the ionization parameter $\xi$ (measured in erg~cm~s$^{-1}$) and the iron abundance $A_{\rm Fe}$ (measured in units of solar iron abundance). In the default flavor {\tt relxill\_nk}, the coronal spectrum is described by a power-law with a high-energy cutoff and we have thus two parameters: the photon index $\Gamma$ and the high-energy cutoff $E_{\rm cut}$. The {\tt Cp} flavors of the model employ the {\tt nthComp} model for the description of the coronal spectrum and the high-energy cutoff $E_{\rm cut}$ is replaced by the coronal temperature $k T_{\rm e}$. The reflection fraction, $R_{\rm f}$, is the relative normalization between the reflection spectrum from the disk and the spectrum of the corona. The latter can be omitted (as in our spectral analysis of GX~339--4 and Swift~J1658.2--4242), and in such a case the radiation from the corona is described by an external model.


\end{document}